\documentclass[preprint,a4paper,number,10pt]{elsarticle}
\usepackage[hidelinks]{hyperref}

\usepackage{amssymb}

\usepackage{lineno}
\usepackage{amsmath,amssymb,amsfonts} 
\usepackage{siunitx}
\usepackage{booktabs}
\usepackage{orcidlink}

\usepackage[export]{adjustbox}
\newcommand{\V}[1]{\boldsymbol{#1}}
\newcommand{\M}[1]{\mathbf{#1}}
\newcommand{\ud}{\mathrm{d}} 							

\let\oldalign\align
\let\oldendalign\endalign

\renewenvironment{align}
  {\linenomathNonumbers\oldalign}
  {\oldendalign\endlinenomath}
\journal{Renewable Energy}

\begin{document}

\begin{frontmatter}



\title{Adjoint~Optimisation for Wind~Farm~Flow~Control with a Free\nobreakdash-Vortex Wake~Model}


\author[1]{Maarten J. van den Broek\orcidlink{0000-0001-7396-4001}\corref{cor1}}
	\ead{m.j.vandenbroek@tudelft.nl}
\cortext[cor1]{Corresponding author}
\affiliation[1]{organization={Delft Centre for Systems and Control, TU Delft},
	addressline={Mekelweg 2},
	postcode={2628 CD},
	city={Delft},
	country={The~Netherlands}
}

\author[2]{Delphine De Tavernier\orcidlink{0000-0002-8678-8198}}
\ead{d.a.m.detavernier@tudelft.nl}
\affiliation[2]{organization={Department of Flow Physics and Technology, Wind Energy, TU Delft},
	addressline={P.O. Box 5058},
	postcode={2600 GB},
	city={Delft},
	country={The~Netherlands}
}

\author[3]{Benjamin Sanderse\orcidlink{0000-0001-9483-1988}}
\ead{b.sanderse@cwi.nl}
\affiliation[3]{organization={Scientific Computing, CWI},
	addressline={P.O. Box 94079},
	postcode={1090 GB},
	city={Amsterdam},
	country={The~Netherlands}
}

\author[1]{{Jan\nobreakdash-Willem} van Wingerden\orcidlink{0000-0003-3061-7442}}
\ead{j.w.vanwingerden@tudelft.nl}



\begin{abstract}
Wind farm flow control aims to improve wind turbine performance by reducing aerodynamic wake interaction between turbines.
Dynamic, physics-based models of wind farm flows have been essential for exploring control strategies such as wake redirection and dynamic induction control.
Free vortex methods can provide a computationally efficient way to model wind turbine wake dynamics for control optimisation.
We present a control-oriented free-vortex wake model of a 2D and 3D actuator disc to represent wind turbine wakes.
The novel derivation of the discrete adjoint equations allows efficient gradient evaluation for gradient-based optimisation
in an economic model-predictive control algorithm.
Initial results are presented for mean power maximisation in a two-turbine case study.
An induction control signal is found using the 2D model that is
roughly periodic and supports previous results on dynamic induction control to stimulate wake mixing.
The 3D model formulation effectively models a curled wake under yaw misalignment.
Under time-varying wind direction, the optimisation 
finds solutions demonstrating both wake steering and a smooth transition to greedy control.
The free-vortex wake model with gradient information shows potential for efficient optimisation and provides a promising way to further explore dynamic wind farm flow control.
\end{abstract}





\end{frontmatter}



\section{Introduction}
Large, densely spaced wind farms are designed and constructed to make use of limited offshore parcels.
Within these farms, aerodynamic interaction between wind turbines reduces power production and increases fatigue loading as turbulent, low-energy wakes travel through the farm and negatively affect downstream turbines.
Wind farm topology is designed to minimise these interactions, but is inflexible to cope with dynamic, varying atmospheric conditions~\citep{VanWingerden2020a}.
The purpose of wind farm control is to minimise the detrimental effects of aerodynamic interaction between wind turbines in a wind farm.

Control strategies for wind farm control can be roughly divided in three categories: wake redirection by yaw misalignment, induction control, and wake mixing strategies~\citep{Meyers2022}.
First, the use of yaw misalignment with respect to the free-stream wind direction to redirect wakes downstream has been shown to effectively improve performance under steady conditions in both wind tunnel experiments~\citep{Campagnolo2016,Bastankhah2019a,Campagnolo2020} and field studies~\citep{Howland2019,Fleming2020,Fleming2021,Doekemeijer2021,Simley2021}.
Second, sinusoidal thrust variations, and consequent induction variations, through collective pitch control have been found to improve wake recovery in an LES study~\citep{Munters2018a} and in wind tunnel experiments~\citep{Frederik2020a}.  
Finally, recent developments in stimulating wake mixing have shown the potential to improve upon collective pitch variations with the helix approach, an individual pitch control strategy~\citep{Frederik2020}.

Control-oriented surrogate models are often at the core of wind farm control algorithms.
Steady-state engineering wake models, such as those that have been implemented in FLORIS~\citep{FLORIS2022}, are the current industry standard.
These include, for example, the Gaussian model~\citep{Bastankhah2014} or a steady representation of curled wake dynamics~\citep{Martinez-Tossas2019}.
As steady-state wake representations are limited in realistic time-varying conditions, dynamic effects have been added to these engineering wake models to improve upon the steady-state results by including dynamic wake meandering~\citep{Larsen2007} or using Lagrangian particles to incorporate wake dynamics~\citep{Becker2022}.

Several studies have also developed physics-based dynamic models for wind farm flow control, especially using the adjoint method to efficiently calculate gradient information for a scalar objective function with a large number of parameters.
The patterns found through optimal control studies with adjoint large-eddy simulations~\citep{Goit2015,Goit2016} provided the basis for dynamic induction control methods, although these simulations are too computationally expensive for real-time control applications~\citep{Munters2018,Munters2018a}.
WFSim provides a 2D Navier-Stokes based wind-farm flow model  for control~\citep{Boersma2018}, which has then been used for adjoint optimisation of induction control~\citep{Vali2019}.
FRED~\citep{FRED2021} builds on the results from WFSim to simulate wind farm performance with the adjoint for gradient calculation~\citep{VandenBroek2020,VandenBroek2022}.
However, the 2D physics inherent in this model lack the curled wake dynamics of a wind turbine under yaw misalignment~\citep{Howland2016,Bastankhah2016,Bartl2018,Fleming2018} and could not accurately model the effects of wake redirection~\citep{VandenBroek2022}.

In contrast to conventional computational fluid dynamic approaches, free-vortex methods use the vorticity formulation of the Navier-Stokes equations to model wind turbine wakes with Lagrangian elements~\citep{Katz2001}.
Within the field of wind energy, free-vortex wake models have been used to study floating wind turbines and wake dynamics~\citep{Lee2019} and to study dynamic wake control methods and analyse wake stability~\citep{Brown2021}.
The latter uses the CACTUS code which has been shown to be mostly accurate for near-wake regions~\citep{Houck2022}.
Even though vortex methods are generally more accurate in near-wake regions, a free-vortex ring method has been used to model far wake dynamics for both fixed-bottom and floating wind turbines~\citep{Dong2019}.
Additionally, an actuator-disc model based on discretised vortex rings has been shown to capture the 3D dynamics of the kidney-shaped wake under yaw misalignment~\citep{Berdowski2018}.

In this paper, we propose the use of the free-vortex wake method
as a computationally efficient, physics-based surrogate wake model for control optimisation, especially coupled with the adjoint for efficient evaluation of the gradient.
This work aims to extend the possibilities for optimisation of induction and yaw signals for dynamic wind farm flow control.
For that purpose, the contribution of this paper is threefold:
(i) a control-oriented free-vortex wake model of an actuator disc in 2D and 3D with the discrete adjoint for gradient computation,
(ii) an economic model-predictive control implementation for dynamic wind farm flow control,
and (iii) initial results that demonstrate dynamic induction control and yaw control under time-varying wind direction.

The remainder of the paper is structured as follows.
A 2D and 3D free-vortex model of an actuator disc to represent a wind turbine wake is presented in Section~\ref{sec:model}.
The non-linear optimisation problem for economic model-predictive control is formulated in Section~\ref{sec:optimisation} together with the discrete adjoint method for calculating the gradient.
Results are discussed in Section~\ref{sec:results}, which provides an overview of operation under steady conditions followed by receding horizon control optimisation of time-varying axial induction and yaw signals.
Finally, conclusions are presented in Section~\ref{sec:conclusions}.





\section{Control-Oriented Free-Vortex Wake Model}\label{sec:model}
The general formulation for the control-oriented free-vortex wake (FVW) representation is described in Section~\ref{subsec:fvw_gen}.
Aspects specific to the 2D and 3D implementations are then defined in Section~\ref{subsec:fvw_2d} and Section~\ref{subsec:fvw_3d}, respectively.
The convergence and validation of the method for the numerical parameters used in this paper is provided in \ref{sec:parameter_choice}.

\subsection{General formulation}\label{subsec:fvw_gen}
An actuator-disc representation of a wind turbine is implemented with the free-vortex method in both a two-dimensional (2D) and three-dimensional (3D) formulation.
The free-vortex method is based on Lagrangian particles that advect downstream.
These particles induce a velocity based on their associated vorticity.
The resultant flow velocity may be calculated at any position based on the free-stream velocity and the sum of induced velocities.
For a further description  of the fundamentals, the reader is referred to aerodynamic literature, such as~\cite{Katz2001}.

The use of the free-vortex wake method requires the assumption of inviscid and incompressible flow.
The actuator disc is assumed to be uniformly loaded so it only releases vorticity along its edge~\citep{Katz2001}.
For the 2D model, the wake is modelled by releasing pairs of vortex points at the edge of the actuator disc at every simulation time-step.
The 3D code is based on the simulation of discretised vortex rings with vortex filaments, adapted from the model described by~\citet{Berdowski2018}.
For convenience, all units have been non-dimensionalised by the rotor diameter and inflow speed.


A system with fixed dimensionality is preferred for control optimisation, therefore the wake models are set up with  $n_\mathrm{e}$ elements per vortex ring and a fixed number of vortex rings $n_\mathrm{r}$.
The number of points to define the vortex elements $n_\mathrm{p}$ equals $n_\mathrm{e}$ in 2D and $n_\mathrm{e}+1$ in 3D.
The spatial dimension of the simulation is $n_\mathrm{d}$, which equals either two or three.
The number of turbines modelled is $n_\mathrm{t}$ and the number of control parameters per turbine is $n_\mathrm{c}$. 
For example, the total number of states is $n_\mathrm{s}=2n_\mathrm{r}n_\mathrm{p}n_\mathrm{d}+n_\mathrm{r}n_\mathrm{e}+n_\mathrm{t}n_\mathrm{c}$ for a single wake modelled with the FVW,
where additional virtual turbines are evaluated using the flow velocity without including their effect on the wake.

We set up the model as a non-linear state-space system in discrete time, 
\begin{align}
	\V{q}_{k+1} &= f(\V{q}_k, \V{m}_k)\,, \label{eq:state_update}\\
		\V{y}_{k} &= g(\V{q}_k, \V{m}_k)\,, \label{eq:state_output}
\end{align}
where for every discrete time step $k$ the updated state $\V{q}_{k+1}\in \mathbb{R}^{n_\mathrm{s}}$ and the output vector $\V{y}_k\in \mathbb{R}^{n_\mathrm{t}}$ are a function of the current state $\V{q}_k \in \mathbb{R}^{n_\mathrm{s}}$ and the control inputs $\V{m}_k\in \mathbb{R}^{n_\mathrm{t}n_\mathrm{c}}$.
The state vector is built up as 
\begin{align}
	\V{q} = \begin{bmatrix}
		\V{X} \\ \V{\Gamma} \\ \V{U} \\ \V{M} 
	\end{bmatrix}\,,
\end{align}
from the vortex element positions $\V{X} \in \mathbb{R}^{n_\mathrm{r}n_\mathrm{p}n_\mathrm{d}}$, the vortex element circulations $\V{\Gamma}\in\mathbb{R}^{n_\mathrm{r}n_\mathrm{e}}$, the stored free-stream velocity $\V{U} \in \mathbb{R}^{n_\mathrm{r}n_\mathrm{p}n_\mathrm{d}}$, and the control inputs from the previous time step $\V{M}\in \mathbb{R}^{n_\mathrm{t}n_\mathrm{c}}$.
The full control vector $\V{m}$ is defined as
\begin{align}
	\V{m} =
	\begin{bmatrix}
		a_0 \\\psi_0 \\a_1 \\\psi_1\\
	\end{bmatrix}\,,
\end{align}
for a two-turbine configuration with axial induction $a$ and turbine yaw angle $\psi$. 

States corresponding to a ring are indicated with a subscript, rings are indexed with a superscript starting from 0.
This allows, for example, the convenient relation of a point $\V{x}_i^{(b)}\in\mathbb{R}^{n_\mathrm{d}}$ to the point in the same position in the previous ring $\V{x}_i^{(b-1)}$,
or all points in a ring $\V{X}^{(b)}\in\mathbb{R}^{n_\mathrm{p}n_\mathrm{d}}$ to all points in the previous ring $\V{X}^{(b-1)}$.

For all rings except the first ($b\geq 1$), the position update is calculated from the position of the previous ring with simulation time step $h$, the stored inflow velocity $\V{u}_{\infty}\in\mathbb{R}^{n_\mathrm{d}}$, and the total induced velocity $\V{u}_\mathrm{ind}\in\mathbb{R}^{n_\mathrm{d}}$,
\begin{align}
	\left.\V{x}_i^{(b)}\right|_{k+1} &= 
	\left.\V{x}_i^{(b-1)} \right|_{k}
	+ \left.h\left(\V{u}_{\infty,i}^{(b-1)} 
	+\V{u}_\mathrm{ind}(\V{x}_i^{(b-1)},\V{q})\right)\right|_{k} \,.
\end{align}
%
The velocity $\V{u}_\mathrm{ind}$ induced at any point $\V{x}$ is the sum of the contribution from all vortex elements in the system
\begin{align}
	\V{u}_\mathrm{ind}(\V{x},\V{q}) = \sum_{b=0}^{n_\mathrm{r}-1}\sum_{j=1}^{n_\mathrm{e}}
	\left.\V{u}_\mathrm{i}\right._{j}^{(b)}\,, 
\end{align}
where $\V{u}_\mathrm{i}\in\mathbb{R}^{n_\mathrm{d}}$ is the velocity induced by a single vortex element.
The generation of new vortex elements in the first ring $\V{X}^{(0)}$ and the velocity induced by a single vortex element $\V{u}_\mathrm{i}$ is defined for 2D and 3D in Section~\ref{subsec:fvw_2d} and Section~\ref{subsec:fvw_3d}, respectively.

The vector $\V{\Gamma}$ contains the vortex strength ${\Gamma}$ for all elements in all rings .
The vortex strength of the first ring is given according to 
%
\begin{align}
\Gamma_i^{(0)}(\V{q}, \V{m}) = \frac{\ud \Gamma}{\ud t} h =c_\mathrm{t}'(a) \frac{1}{2} (\V{u}_{\text{r}}\cdot\V{n}(\psi))^2 h
\quad \text{for } i=1,2,\dots,n_{\mathrm{e}} \,.
\end{align}
In this expression, $\V{u}_\text{r}$ is the average wind speed at the rotor.
The vector $\V{n}\in\mathbb{R}^{n_\mathrm{d}}$ is  a unit vector orthogonal to the rotor disc, pointing in the downstream direction, with the rotation matrix $\M{R}_z\in\mathbb{R}^{n_\mathrm{d}\times n_\mathrm{d}}$ and axis-aligned unit vector $\V{e}_x\in\mathbb{R}^{n_\mathrm{d}}$,
\begin{align}
	\V{n}(\psi) = \M{R}_z(\psi) \V{e}_x \,.
\end{align}
The local thrust coefficient $c_\mathrm{t}'$, is calculated from the axial induction $a$ as
\begin{align}
	c_\mathrm{t}'(a) = \left\{\begin{array}{ll}
	\frac{4a(1-a)}{(1-a)^2} = \frac{4a}{1-a} 	& \text{if } a\leq a_\mathrm{t}\,,\\
	\frac{c_\mathrm{t1}-4(\sqrt{c_\mathrm{t1}}-1)(1-a)}{(1-a)^2} 
	& \text{if } a>a_\mathrm{t}\,,
	\end{array}
	\right.
\end{align}
where the induction $a_\mathrm{t}$ at the transition point is
\begin{align}
	a_\mathrm{t} = 1 - \frac{1}{2} \sqrt{c_\mathrm{t1}}\,,
\end{align}
and the parameter $c_\mathrm{t1}=2.3$.
The thrust coefficient calculation is based on momentum theory with a transition to a linear approximation for high induction values that is an empirical correction based on the Glauert correction~\citep{Burton2001}.
%
Vortex strength of subsequent rings is inherited downstream,
\begin{align}
\left.\Gamma_i^{(b)}\right|_{k+1} = \left.\Gamma_i^{(b-1)}\right|_{k}
\quad \text{for } i=1,2,\dots,n_{\mathrm{e}} \text{ and } b=1,2,\dots,n_\mathrm{r}-1 \,.
\end{align}

Ring zero is initialised at the turbine position with the free-stream velocity, which may vary over space and simulation time,
\begin{align}
	\left.
	\V{U}^{(0)}
	\right|_{k+1} = \V{u}_{\infty}(\V{X}^{(0)},k) \,.
\end{align}
The inflow velocity is then propagated downstream with the state update
\begin{align}
\left.\V{U}^{(b)}\right|_{k+1} = \left.\V{U}^{(b-1)}\right|_{k}
\quad \text{for } b=1,2,\dots,n_\mathrm{r}-1 \,.
\end{align}

The vector $\V{M}$ is an augmentation of the system state to store controls for power calculation at the next time-step,
\begin{align}
	\left.\V{M}\right|_{k+1} = \V{m}_k\,.
\end{align}
This avoids a direct feed-through of control actions to the output function.

The output vector $\V{y}$ contains the power of all turbines as
\begin{align}
	\V{y} = \begin{bmatrix}
		P_0 \\ P_1
	\end{bmatrix}\,, 
\end{align}
for a two-turbine case.
The power $P$ at turbine $i$ is calculated as
\begin{align}
	P_i = \frac{1}{2}c_\mathrm{p}'(a) A_\mathrm{r} (\V{u}_\mathrm{r}\cdot \V{n}(\psi))^3\,, 
\end{align}
with the local power coefficient $c_\mathrm{p}'$, rotor area $A_\mathrm{r}$, the disc-averaged velocity $\V{u}_\mathrm{r}\in\mathbb{R}^{n_\mathrm{d}}$, and the yaw angle $\psi$.
The local power coefficient is   calculated with the induction factor $a$ as
\begin{align}
	c_\mathrm{p}'(a) = 
	\frac{4a(1-a)^2}{(1-a)^3} 
	= \frac{4a}{1-a}\,.
\end{align}
For the disc-averaged velocity, we distribute $n_\mathrm{u}$ points over a disc representing the turbine according to an equal-area distribution~\citep{Masset2011} and rotate the disc over the yaw angle.
The rotor-disc averaged velocity is then
\begin{align}
	\V{u}_\mathrm{r} =  \frac{1}{n_\mathrm{u}}\sum_{i=1}^{n_\mathrm{u}}\left(
	\V{u}_\infty(\V{x}_i,\V{q}) +
	 \V{u}_\mathrm{ind}(\V{x}_i,\V{q})\right)\,,
\end{align}
where the local free-stream flow is calculated as an average from neighbouring points weighted by distance,
\begin{align}
	\V{u}_\infty(\V{x},\V{q}) =  \sum_{i=0}^{n_\mathrm{p}} \sum_{b=0}^{n_\mathrm{r}} \bar{w}_i^{(b)} \V{u}_{\infty,i}^{(b)}\,,
\end{align}
with normalised weights $\bar{w}_i^{(b)}$,
\begin{align}
	w_i^{(b)} &= 
	{\exp(-10||\V{x}-\V{x}_i^{(b)}||)} \,,\\
	\bar{w}_i^{(b)} &= \frac{{w_i}^{(b)}}
	{\sum_{i=0}^{n_\mathrm{p}} \sum_{b=0}^{n_\mathrm{r}-1} w_i^{(b)}} \,.
\end{align}
%
For calculation of power of a virtual turbine -- one that does not act on the flow simulation, but is included for purposes of optimisation -- we lower the  disc-averaged velocity by the induction factor
\begin{align}
	\V{u}_\mathrm{r}^* = (1-a)\V{u}_\mathrm{r} \,.
\end{align}



\subsection{Two-dimensional model specifics}\label{subsec:fvw_2d}
The $n_\mathrm{e}=2$ vortex elements of the first ring are initiated at the edge of the rotor disc with radius $r$
\begin{align}
\left.\V{x}_0^{(0)}(\psi)\right|_{k+1}
 = \M{R}_z(\psi_k)\begin{bmatrix}
0 \\ r
\end{bmatrix}\,,\quad
\left.\V{x}_1^{(0)}(\psi)\right|_{k+1}
 = \M{R}_z(\psi_k)\begin{bmatrix}
0 \\ -r
\end{bmatrix}\,,
\end{align}
where $\M{R}_z(\psi)$ is the rotation matrix for a rotation of an angle $\psi$ around the $z$-axis,
\begin{align}
	\M{R}_z(\psi) = \begin{bmatrix}
		\cos\psi & \sin\psi \\
		-\sin \psi & \cos\psi \\
	\end{bmatrix}\,.
\end{align}
The velocity $\V{u}_\mathrm{i}$ induced at point $\V{x}_0$ by a single vortex element located at $\V{x}_1$ in 2D is calculated with the Biot-Savart law as 
%
\begin{align}
	\V{u}_\mathrm{i}(\V{x}_0, \V{x}_1) = 
\begin{bmatrix}-r_y\\r_x\end{bmatrix}
\left(
\frac{\Gamma}{2\pi}
\frac{1}{||\V{r}||^2}
\right)
\left(
1 - \exp\left(-\frac{||\V{r}||^2}{\sigma^2}\right)
\right)\,,
\label{eq:ui_2d}
\end{align}
where the relative position $\V{r}$ is
\begin{align}
	\V{r} = \V{x}_1 - \V{x}_0\,.
\end{align}
A Gaussian core with core size $\sigma$ is included to regularise singular behaviour of the induced velocity close to the vortex element.

\subsection{Three-dimensional model specifics}\label{subsec:fvw_3d}
At every time-step, the vortex filaments that make up a new vortex ring discretised with $n_\mathrm{e}$ elements are distributed over a circle with radius $r$, with yaw angle $\psi$,
\begin{align}
\left.\V{x}_i^{(0)}(\psi)\right|_{k+1}
 = \M{R}_z(\psi_k)\begin{bmatrix}
0 \\ r\cos(2\pi \frac{i}{n_\mathrm{e}}) \\ r\sin(2\pi \frac{i}{n_\mathrm{e}})
\end{bmatrix} \quad \text{for } i={0,1,\dots,n_\mathrm{e}} \,,
\end{align}
where $\M{R}_z(\psi)$ is the rotation matrix for a rotation of an angle $\psi$ around the $z$-axis,
\begin{align}
	\M{R}_z(\psi) = \begin{bmatrix}
		\cos\psi & \sin\psi & 0 \\
		-\sin \psi & \cos\psi & 0\\
		0 & 0 & 1\\
	\end{bmatrix}\,.
\end{align}
The induced velocity $\V{u}_\mathrm{i}$ at a point $\V{x}_0$ is calculated with with Biot-Savart law from a single vortex element starting at $\V{x}_1$ and ending at $\V{x}_2$, with vortex strength $\Gamma$,
\begin{align}
	\V{u}_{\mathrm{i}}(\V{x}_0, \V{x}_1, \V{x}_2) = 
\left(
\frac{\Gamma}{4\pi}
\frac{\V{r}_{1}\times \V{r}_{2}}{||\V{r}_{1}\times\V{r}_{2}||^2}
\right)
\left(
\V{r}_{0}\cdot
\left(
\frac{\V{r}_{1}}{||\V{r}_{1}||}
- \frac{\V{r}_{2}}{||\V{r}_{2}||}
\right)
\right)
\left(
1 - \exp\left(-\frac{||\V{r}_1 \times \V{r_2}||^2}{\sigma^2||\V{r}_0||^2}\right)
\right)\,,
\end{align}
where the relative positions $\V{r}$ are defined as
\begin{align}
\V{r}_{0} &= \V{x}_2 - \V{x}_1\,,\\
\V{r}_{1} &= \V{x}_1 - \V{x}_0\,,\\
\V{r}_{2} &= \V{x}_2 - \V{x}_0\,.
\label{eq:ui_3d}
\end{align}
A Gaussian core with core size $\sigma$ is included to regularise singular behaviour of the induced velocity close to the vortex filament.


\section{Optimisation for Power Maximisation}\label{sec:optimisation}
The free-vortex wake model described in the previous section is implemented as a novel surrogate model for dynamic wind farm flow control.
Wind turbine power maximisation is introduced in Section~\ref{subsec:empc} in an economic model-predictive control setting.
The associated non-linear optimisation problem is formulated in Section~\ref{subsec:objective}.
The derivation of the discrete adjoint for calculation of the gradient is described in 
Section~\ref{subsec:adjoint}, followed by the choice of a gradient-based optimisation method to solve the non-linear problem in Section~\ref{subsec:optimisation}.

\subsection{Economic model-predictive control}\label{subsec:empc}
The conventional model-predictive control (MPC) approach is a model-based optimisation of control signals to drive an objective functional to zero, for example for optimal tracking of a reference signal.
However, for maximisation of wind farm power production, the optimal objective value is not known a priori, leading to an economic problem formulation.
The economic MPC (EMPC) approach considers optimisation of an objective to an unknown extremum.
This optimisation problem is conventionally solved in a receding horizon setting  with a finite prediction horizon.
After optimisation, the first (set of) control(s) is implemented and the problem is shifted and solved again up to the new horizon~\citep{Grune2017}.

One problem with optimisation to a finite horizon is that the optimisation considers the prediction horizon as the end of time.
Therefore, control actions that prioritise gain within the horizon may be optimal, although they would have undesired consequences post-horizon.
This is known as the turnpike effect~\citep{Dorfman1958}, where a solution stays close to the optimal trajectory for most of the window but diverges towards the horizon.
These finite horizon effects may be treated by terminal constraints or terminal conditions~\citep{Grune2017}. 
For example, the control signal has been kept constant towards the horizon to limit undesired effects in wind farm control~\citep{Vali2019} or a terminal condition on rotor kinetic energy has been used to regularise optimisation results for wind turbine control~\citep{Gros2013}.
Given a sufficiently long prediction horizon, EMPC has been shown to also converge without terminal constraints~\citep{Grune2013}. 

In this paper, the turnpike effects are treated by considering sufficiently long prediction horizons within the receding horizon setting, so as not to require terminal constraints.
The control problem is formulated in a non-linear EMPC setting without terminal conditions with the goal of maximising mean power production over time.

\subsection{Objective function definition}\label{subsec:objective}
A non-linear minimisation problem with a scalar objective function $J$ is constructed to find the set of optimal controls $\V{m}_{k_0+i}\in\mathbb{R}^{n_\mathrm{m}}$, with $n_\mathrm{m}\leq n_\mathrm{t}n_\mathrm{c}$ the number of free controls and $i=0,1,\dots,N_\mathrm{h}$. 
The objective is the total power output over a prediction horizon of $N_\mathrm{h}$ steps from the current step $k_0$,
\begin{align}
\label{eq:objective}
  \min_{\V{m}_k} \sum_{k=k_0}^{k_0+N_\mathrm{h}} J(\V{q}_k,\V{m}_k) 
  = \min_{\V{m}_k} \sum_{k=k_0}^{k_0+N_\mathrm{h}} \M{Q}\V{y}_k(\V{q}_k,\V{m}_k) + \V{\Delta m}_k^{\operatorname{T}}\M{R}\V{\Delta m}_k
    \,,
\end{align}
where $\V{y}_k$ contains the power of modelled and virtual turbines, $\V{\Delta m}_k = \V{m}_k-\V{m}_{k-1}$ is the change in control value between time steps, and $\M{Q}\in\mathbb{R}^{1\times n_\mathrm{t}}$ and $\M{R}\in\mathbb{R}^{n_\mathrm{m}\times n_\mathrm{m}}$ are weights to balance power output and actuation cost.
%
A linear sum of power is chosen for mean power maximisation because power is already 
a positive objective function.
A quadratic functional would more heavily weight peaks in power production and be suboptimal for maximisation of mean power.
The output weight is chosen negative ($\M{Q}<0$) so that power is maximised for minimisation of the objective.
The input weight $\M{R}$ functions as a regularisation term and aids convergence to suitable control solutions by smoothing the optimisation landscape.

\subsection{Discrete adjoint method for constructing the gradient}\label{subsec:adjoint}
The gradient of the objective function is calculated following the discrete adjoint method~\citep{Lauss2017}
because the method scales well for a large number of input sensitivities.
We take the non-linear state-space system in \eqref{eq:state_update} and define the objective function $J_k=J(\V{q}_k, \V{m}_k)$ at time-step $k$,
such that the total objective function $J_\mathrm{total}$ is accumulated over a number of steps  $N_\mathrm{h}$,
\begin{align}
	J_\mathrm{total}= J_{N_\mathrm{h}} + \sum_{i=0}^{N_\mathrm{h}-1} J_i\,,
\end{align}
where $i=0$ at the current time-step $k=k_0$.
This is the total objective function to be minimised in the optimisation problem in~\eqref{eq:objective}.

To derive the adjoint system, we extend the objective function with adjoint states and system constraint,
\begin{align}
	\bar{J}_\mathrm{total} = J_{N_\mathrm{h}} 
		+ \sum_{i=0}^{{N_\mathrm{h}} -1}\left(
				 J_i + \V{\lambda}_{i+1}^\mathrm{T}\left(
				 	f_i - \V{q}_{i+1}
				 \right)
			\right)\,,
\end{align}
where the adjoint states $\V{\lambda}$ can be chosen freely because $f_i-\V{q}_{i+1}=0$.
%
Since $J_k=J(\V{q}_k,\V{m}_k)$, a differential change $\delta \bar{J}_\mathrm{total}$ can be expanded in terms of changes in $\V{q}$ and $\V{m}$ as:
\begin{align}
\label{eq:variations}
	\delta \bar{J}_\mathrm{total} = &\left(\frac{\partial J_{N_\mathrm{h}}}{\partial \V{q}_{N_\mathrm{h}}} - \V{\lambda}_{N_\mathrm{h}}^\mathrm{T} \right)\delta \V{q}_{N_\mathrm{h}} + \frac{\partial J_{N_\mathrm{h}}}{\partial \V{m}_{N_\mathrm{h}}} \delta \V{m}_{N_\mathrm{h}} \nonumber\\
	&+ \sum_{i=0}^{{N_\mathrm{h}}-1}\left(
				\left(\frac{\partial J_i}{\partial \V{q}_i} + \V{\lambda}_{i+1}^\mathrm{T}\frac{\partial f_i}{\partial \V{q}_i} - \V{\lambda}_i^\mathrm{T} \right) \delta \V{q}_i + 
				\left(\frac{\partial J_i}{\partial \V{m}_i} + \V{\lambda}_{i+1}^\mathrm{T}\frac{\partial f_i}{\partial \V{m}_i}\right) \delta \V{m}_i
			\right) \,,
\end{align}
We then choose the adjoint states to be
\begin{align}
\label{eq:adjointstates}
	\V{\lambda}_{N_\mathrm{h}}^\mathrm{T} =  \frac{\partial J_{N_\mathrm{h}}}{\partial \V{q}_{N_\mathrm{h}}}\,, \qquad
	\V{\lambda}_i^\mathrm{T} = \frac{\partial J_i}{\partial \V{q}_i} + \V{\lambda}_{i+1}^\mathrm{T}\frac{\partial f_i}{\partial \V{q}_i} \,,
	\qquad \V{\lambda}_0^\mathrm{T}=\V{0}\,,
\end{align}
such that the variations 
due to $\V{q}$  in \eqref{eq:variations} are cancelled out.
The adjoint states are solved for by propagation backwards in time, starting from the final adjoint state.
The gradient of the objective function parts $J_k$ to the input can then be calculated from these adjoint states
\begin{align}
\nabla J_{N_\mathrm{h}} =  \frac{\delta J_{N_\mathrm{h}}}{\delta \V{m}_{N_\mathrm{h}}} \,, \quad
	\nabla J_i = \frac{\partial J_i}{\partial \V{m}_i}
		 + \left(\frac{\partial f_i}{\partial \V{m}_i}\right)^\mathrm{T} \V{\lambda}_{i+1} \,.
\end{align}
The total gradient $\nabla \bar{J}_\mathrm{total}$ with respect to all control parameters $\V{m}_i$ is then constructed as
\begin{align}
	\nabla \bar{J}_\mathrm{total} =
	\sum_{i=0}^{N_\mathrm{h}}\nabla J_i \,.
\end{align}
The partial derivatives of the state update and output function with respect to the full model state and controls are stored in memory during the forward simulation of the model.
These partial derivatives are provided for the given model and objective function in \ref{sec:derivatives}.

The evaluation of the gradient thus requires a single forward simulation with evaluation of the partial derivatives and a single backward pass to solve for the adjoint states and construct the gradient.
In that sense, this method of gradient evaluation is considerably more efficient than finite difference methods as the computational cost of the discrete adjoint increases only minimally with the number of control parameters for which the derivative is required.
The computational cost of gradient evaluation with the discrete adjoint primarily scales with the expense of the forward simulation and the associated partial derivatives.

\subsection{Gradient-based optimisation methods}\label{subsec:optimisation}
The availability of the gradient allows the use of gradient-based optimisation techniques for control optimisation.
Exploration of the objective function shows that it is non-linear and non-convex, with almost flat regions and numerous local minima.
Initial experiments were run with L-BFGS-B optimisation \citep{Byrd1995} as was also used in the work by \citet{Munters2018}.
However, this optimiser appeared sensitive to initialisation at local maxima and to convergence to local minima.

The Adam optimiser \citep{Kingma2015} is a gradient-based method  often used in machine learning for optimisation of neural network weights, where it is applied for gradient descent with noisy gradients in complex optimisation landscapes.
It uses a momentum approach to accelerate gradient descent and has proven to be less sensitive to the choice of initial guess and local minima.
Within this work, we use the Adam optimiser with the default parameters; a maximum step size $\alpha=0.001$ and the default decay rates $\beta_1=0.9$ and $\beta_2=0.999$.
Tuning of these parameters may still improve performance.
The yaw angle is on a different order of magnitude than axial induction.
Therefore, it is scaled by a factor $10^{-2}$ in the optimisation, so that the step size covers a similar range of the allowable range of induction value and yaw angle.




\section{Results and Discussion}\label{sec:results}
A brief overview of the 2D and 3D FVW under steady conditions is given in Section~\ref{sub:steady} to illustrate the test case configuration and provide a steady baseline for control performance.
This is followed by two example cases to demonstrate the use of the FVW as a novel surrogate model for control optimisation 
in the receding horizon setting described previously;
a 2D case with induction control is provided in Section~\ref{subsec:induction_results} and a 3D case for yaw control under time-varying wind direction in Section~\ref{subsec:yaw_results}.
Finally, Section~\ref{subsec:empc_discussion} discusses
finite horizon effects in EMPC for wind farm control.

\subsection{Steady state operation}\label{sub:steady}
We define a two-turbine case for evaluating control optimisation with the FVW, starting with steady-state control characteristics.
The two turbines are spaced \SI{5}{D} apart, where $D$ is the rotor diameter, aligned with the uniform unit inflow.
The upstream turbine is modelled with the FVW and the virtual downstream turbine performance is evaluated using the flow velocity over the rotor area at the downstream position.
The parameters for the FVW are provided in Table~\ref{tab:par_2d}.
{An exploration of parameter sensitivity is supplied in \ref{sec:parameter_choice}}.
\begin{table}[!hbt]
\centering
\caption{FVW parameters for 2D and 3D case}\label{tab:par_2d}
	\begin{tabular}{l c c c}
			\toprule
								&					& 2D & 	{3D} \\
			\midrule
			time-step            & $h$               & 0.2   & 0.3\\
			core size            & $\sigma$          & 0.1   & 0.16\\
			number of rings      & $n_\mathrm{r}$    & 60    & {40}\\
			elements per ring    & $n_\mathrm{e}$    & 2     & {16}\\
			\bottomrule
	\end{tabular}
\end{table}

\begin{figure}[!b]
	\centering
	\includegraphics[width=\linewidth]{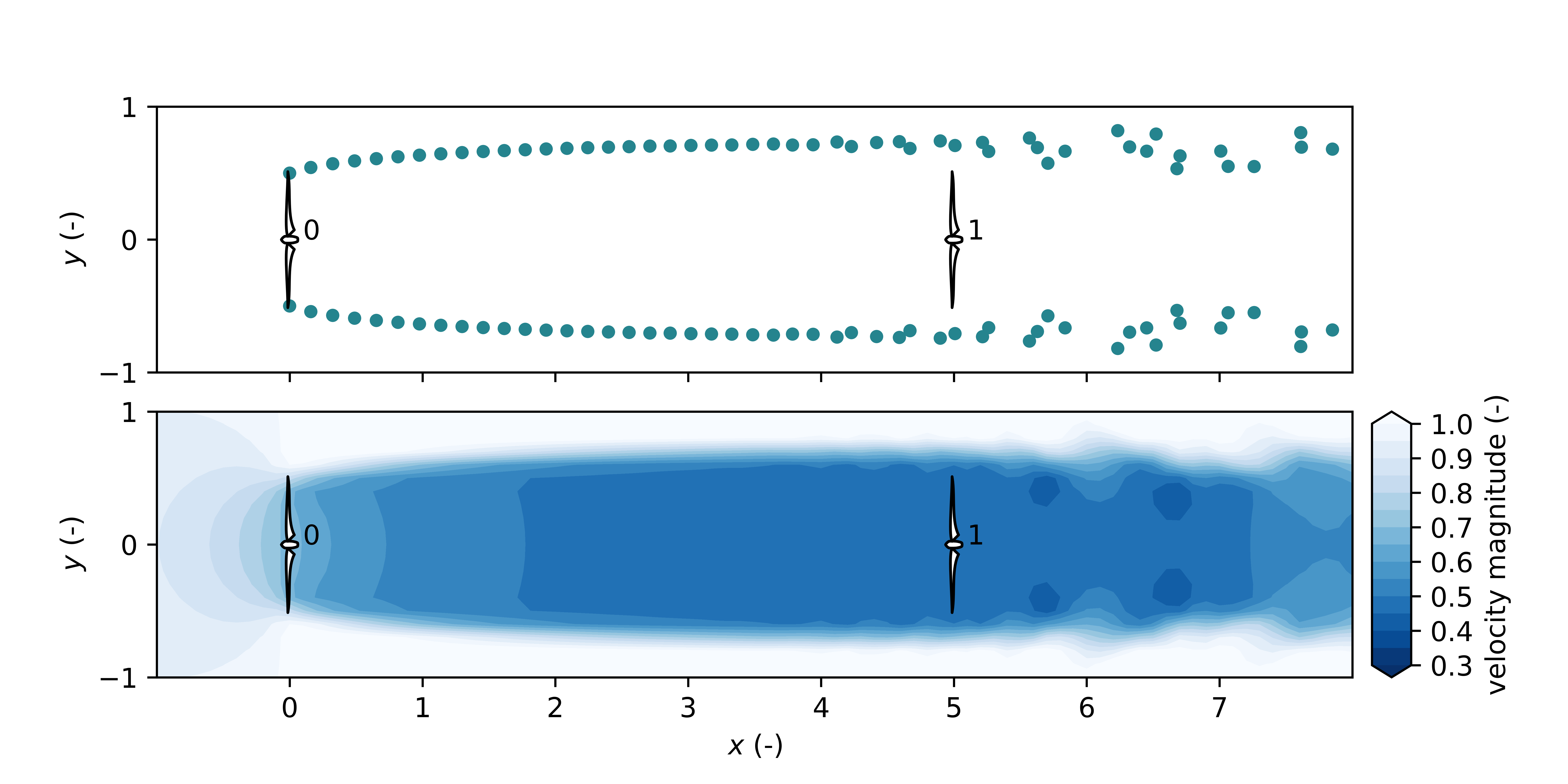}
	\caption{Illustration of the 2D FVW for uniform unit inflow without yaw misalignment and with a constant induction factor $a=0.33$.
	The pairs of vortex points (top) can be used to calculate the velocity at any point, allowing visualisation of a dense velocity field (bottom).
	The figure illustrates the two-turbine case where the second turbine performance is calculated from the flow velocity \SI{5}{D} downstream.
	}
	\label{fig:flow_2d_steady}
\end{figure}

Figure~\ref{fig:flow_2d_steady} shows a 2D FVW simulation with an induction factor $a=0.33$ and without yaw misalignment.
The pairs of vortex points provide the basis for the simulation and allow calculation of a dense velocity field.
Disturbances in the far wake are the result of numerical instabilities.
It is also notable that the wake is quite wide as is expected for planar flow.

\begin{figure}[!b]
	\centering
    \includegraphics[width=1.3\linewidth,center]{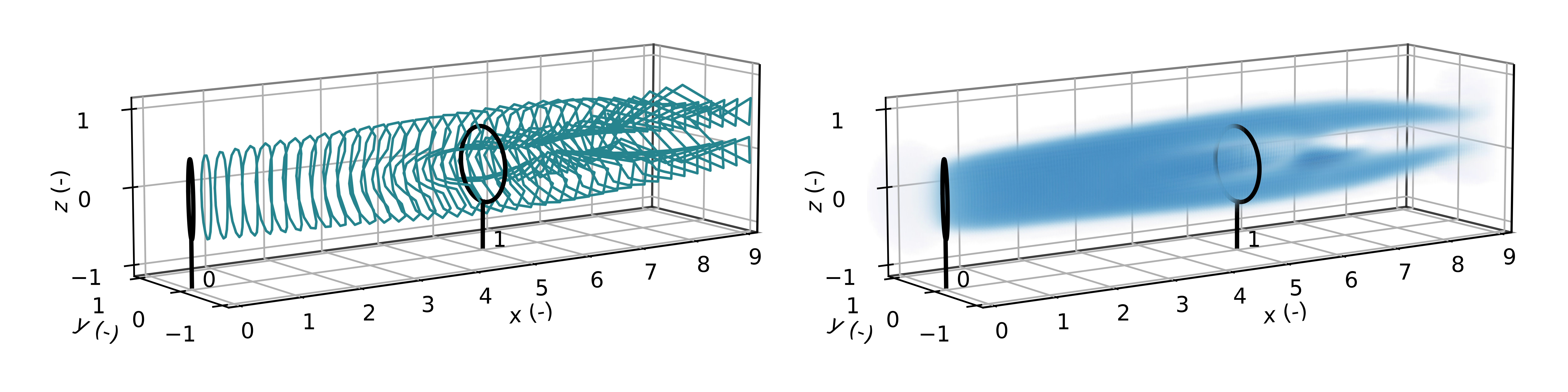}
    \caption{
    The 3D FVW models the wake from a series of vortex rings discretised into vortex elements (left), allowing calculation of a velocity field showing the wake deficit (right).
	{The kidney-shaped wake appears as a pair of counter-rotating vortices is formed under yaw misalignment.}
    Simulation under uniform inflow with a yaw misalignment of {$\psi=\SI{30}{\degree}$} and induction $a=0.33$.
    The figure illustrates the case where the upstream turbine is modelled with the 3D FVW and the downstream turbine performance is calculated from the flow velocity over a rotor disc $\SI{5}{D}$ downstream.
	}
	\label{fig:FVW3D}
\end{figure}

The 3D FVW produces a vortex ring structure as shown in Figure~\ref{fig:FVW3D} for a simulation with yaw misalignment of {$\psi=\SI{30}{\degree}$} and induction factor $a=0.33$.
The figure shows the dense velocity field with the wake deficit calculated from the skeleton of vortex filaments.
A kidney-shaped wake appears from the pair of counter-rotating vortices that are generated by a turbine operating under yaw misalignment, as shown in \citep{Berdowski2018}.

\begin{figure}[!b]
	\centering
	\includegraphics[width=\textwidth]{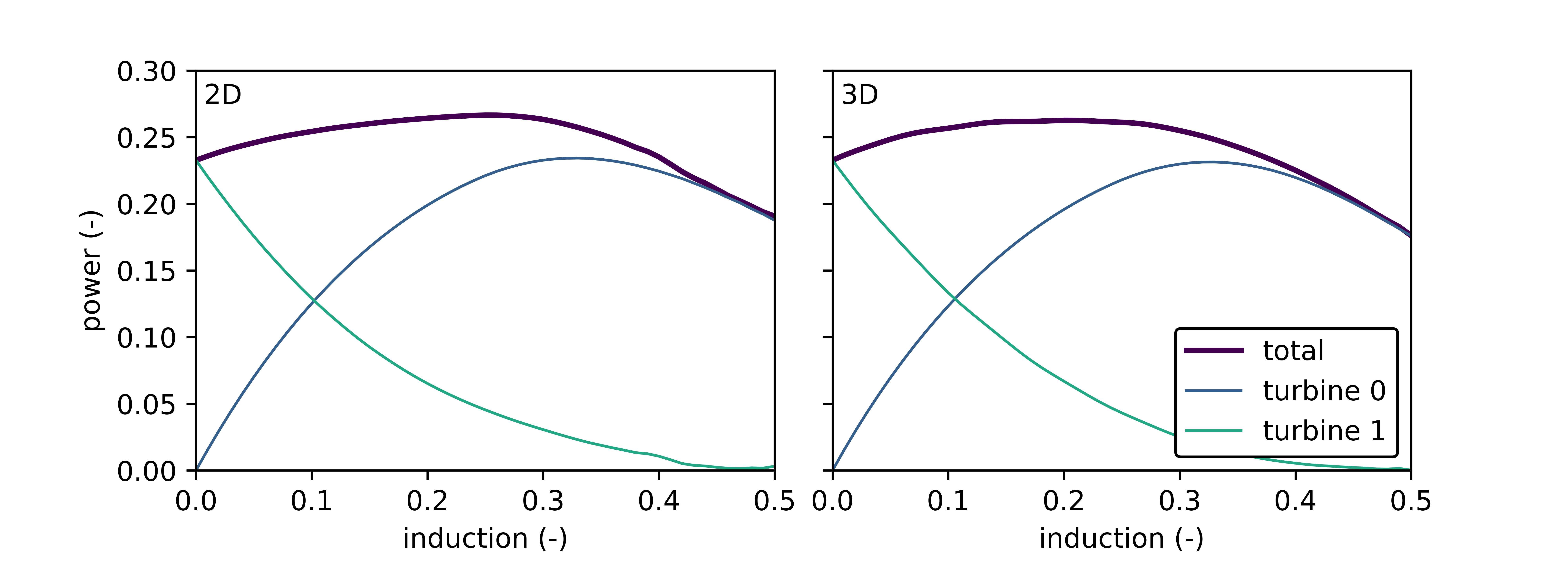}
	\caption{Power production in steady state for varying induction on the upstream turbine, in 2D (left) and 3D (right), with turbine configuration as in Figures~\ref{fig:flow_2d_steady}~and~\ref{fig:FVW3D} respectively.
	Total power is the sum of power from turbine 0 and turbine 1.
	Maximum greedy power production occurs for $a=0.33$.
	Within this model, lowering the induction on the upstream turbine to {$a=0.27$} leads to a {\SI{3.6}{\%}} gain in total power.}
	\label{fig:inductionsweep}
\end{figure}

The model response to control signal variation is verified by examining power production in steady state. First, Figure~\ref{fig:inductionsweep} shows the 2D and 3D FVW power curve for a variation in axial induction from $a=0$ to $a=0.5$.
The maximum individual turbine power matches the expectation from momentum theory for the chosen parameters at the theoretical optimum induction of $a=0.33$.
Steady under-induction provides a power gain of {\SI{3.6}{\%}} over greedy control.
The 2D and 3D FVW show remarkably similar behaviour in terms of power production for varying induction factor on the upstream turbine.
The similarity in the power estimate shows an opportunity for doing induction control in 2D model studies.
Additionally, 2D wind farm flow models have already been used for studies of induction control in a wind farm setting~\citep{VanWingerden2017,Vali2019}.

\begin{figure}[!t]
	\centering
	\includegraphics[width=\textwidth]{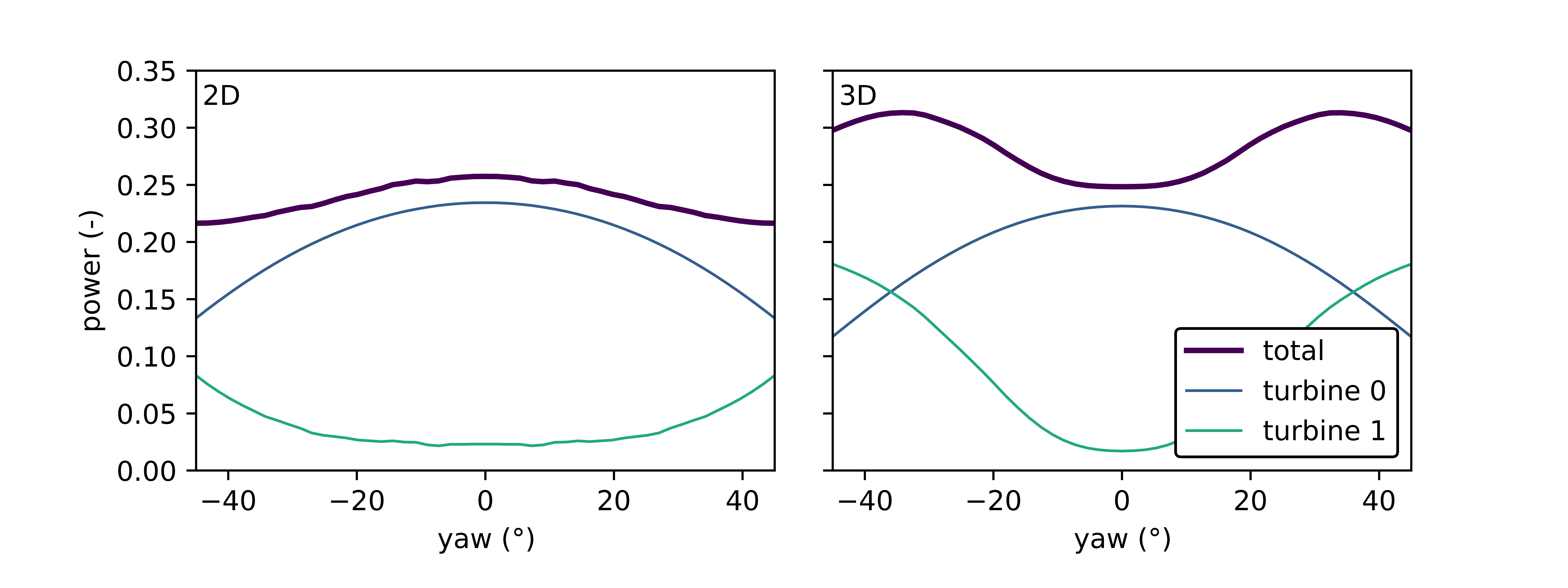}
	\caption{Power production in steady state for varying yaw on the upstream turbine, in 2D (left) and 3D (right), with turbine configuration as in Figures~\ref{fig:flow_2d_steady}~and~\ref{fig:FVW3D} respectively.
	The downstream turbine power is calculated from the flow velocity \SI{5}{D} downstream from the first turbine.
	Total power is the sum of power from turbine 0 and turbine 1.
	The 2D FVW does not have the dynamics to model wake steering effectively.
	In the 3D FVW, a {\SI{26.1}{\%}} gain in total power is observed for a yaw misalignment angle of {$\psi=\SI{34}{\degree}$} compared to greedy control where $\psi=\SI{0}{\degree}$.
	}
	\label{fig:yawsweep}
\end{figure}

Second, a yaw sweep from $\psi=\SI{-45}{\degree}$ to $\psi=\SI{45}{\degree}$ is illustrated in Figure~\ref{fig:yawsweep}.
This steady sweep shows a demonstrable lack of power gain from yaw misalignment in the 2D FVW. 
However, in 3D, yaw misalignment does lead to wake redirection and maximum power achieved in steady-state is {\SI{0.313}{}} for a misalignment angle of {\SI{34}{\degree}}, providing a gain of  {\SI{26.1}{\%}} over greedy control.

The 3D FVW shows the formation of a kidney-shaped wake from a counter-rotating vortex pair when the turbine is operated under yaw misalignment.
The subsequent deflection of the turbine wake leads to an increase of the combined power production.
These dynamics are not present in 2D, which may explain the lack of wake redirection.
This supports previous results that found 2D flow modelling ineffective in capturing the essential effects of wake steering~\citep{VandenBroek2022}.

The model is currently symmetric, which means there is no difference between positive or negative yaw misalignment.
Experimental studies have found wake steering to be asymmetric due to the rotation induced by the rotor~\citep{Bastankhah2016,Bartl2018,Fleming2018}.
A normal actuator disc was chosen for simplicity, but a root vortex could be included to model the turbine as a rotating actuator disc to model the asymmetric aspect of wake redirection.

\subsection{Induction control in two-dimensional flow}\label{subsec:induction_results}
Given the similarity between 2D and 3D in power curves for variation of axial induction, an optimisation case for induction control is set up in 2D with a configuration as in Figure~\ref{fig:flow_2d_steady} and parameters as in Table~\ref{tab:par_2d}.
Both turbines are aligned with the inflow wind direction and set to a $\psi=\SI{0}{\degree}$ yaw angle.
The downstream turbine is assumed to be performing at its greedy optimum with an induction factor $a=0.33$, whereas the induction control signal of the upstream turbine is to be found by solving the optimisation problem.

The objective function \eqref{eq:objective} is constructed with the control signal $\V{m}_k=\begin{bmatrix}
	a_k
\end{bmatrix}$, over a prediction horizon of $N_\mathrm{h}=100$ samples.
The output contains the power of both turbines, $\V{y}_k=\begin{bmatrix}
	p_0 & p_1
\end{bmatrix}^\mathrm{T}$.
The objective function weights are set to {$\M{Q}=\begin{bmatrix} -1 & -1 \end{bmatrix}$ and $\M{R}=\begin{bmatrix} 10 \end{bmatrix}$}.
This choice of input weight resulted in an adequate balance between input action and power production in an exploratory parameter sweep.
At every time step, the optimisation is run for 50 iterations, 
after which the first value of the control signal from the optimisation is implemented in the receding horizon control scheme.
Further iterations have not lead to consistently better performance in terms of objective function value given the current optimiser configuration.
The starting state for the optimisation case is the result of a steady simulation under greedy operating conditions to remove transient effects.
This initial condition is shown in the first frame of Figure~\ref{fig:vw2d}.

\begin{figure}[!ht]
\centering
\includegraphics[width=1.3\linewidth,center]{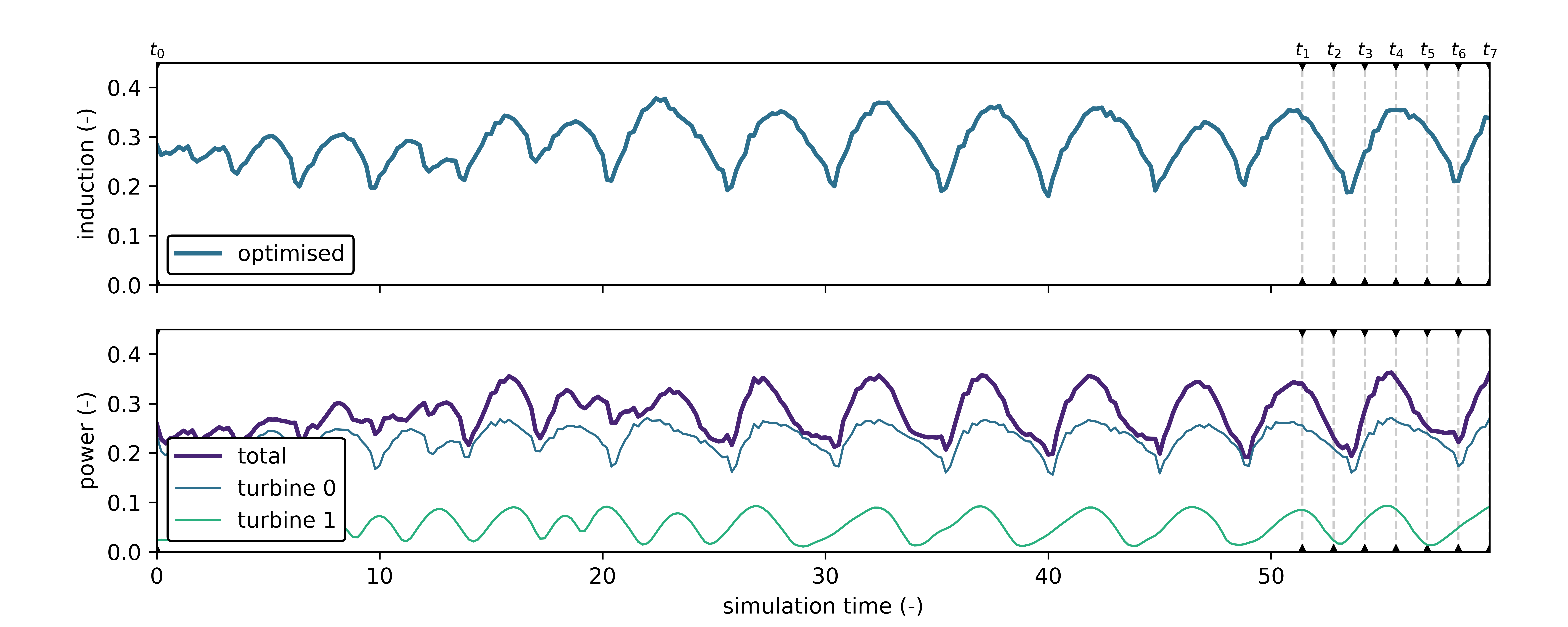}
\caption{
Receding horizon optimisation of the axial induction control signal for total power yields dynamic behaviour that stimulates wake breakdown and improves time-average power production by \SI{6.0}{\%} over steady under-induction.
Control signal for turbine 0 is optimised whilst turbine 1 is virtually modelled to be operating at its greedy optimum, positioned \SI{5}{D} downstream in fully waked conditions.
Total power is the sum of power from turbine 0 and turbine 1.
Snapshots of the flow field at times $t_0$ to $t_7$ are illustrated in Figure~\ref{fig:vw2d}.
}
\label{fig:DIC}
\end{figure}

\begin{figure}[!ht]
\centering
\includegraphics[width=1.3\linewidth,center]{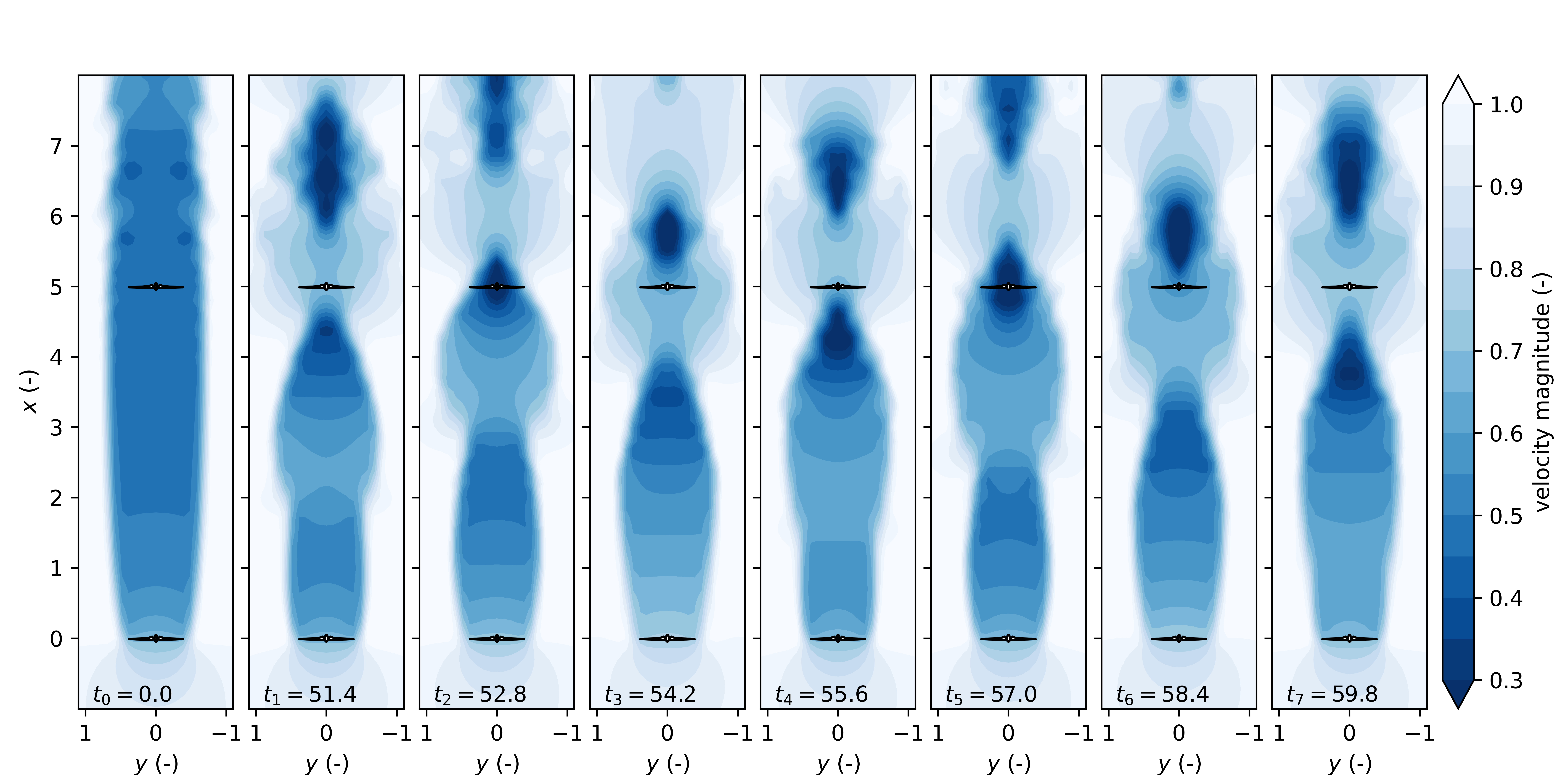}
\caption{
Snapshots of the flow field from the 2D free-vortex wake simulation  under uniform inflow in positive $x$ direction with the induction signal applied to the turbine as shown in Figure~\ref{fig:DIC}.
The effects of the periodic induction excitation can be seen in the structure of the wake and appear to enhance wake breakdown.
}
\label{fig:vw2d}  
\end{figure}

The control signal produced in this economic MPC framework is illustrated in Figure~\ref{fig:DIC}.
The optimisation converges to a roughly periodic excitation with a dominant frequency of approximately {$f=0.20$}.
The signal features sharp downward peaks where the induction is lowered, thus reducing thrust and allowing more flow to pass through the rotor disc.
In addition to the periodic excitation, the mean induction factor is lowered to {$\bar{a}=0.30$} below the greedy optimum of $a=0.33$.
The mean power produced in the final two-thirds of the simulation ($t>20$) is {\SI{0.283}{}}, which is an increase of {\SI{6.0}{\%}} over the maximum power achieved with steady induction control.

A series of snapshots of the wake produced with this control signal is shown in Figure~\ref{fig:vw2d}.
The effects of periodic induction excitation are apparent in the flow field of the wake as coherent structures are formed that travel downstream.

We observe that optimisation
with the 2D FVW leads to induction control signals that combine static under-induction with a strong dynamic component.
Within the current model, this combination of periodic excitation and lowering of the mean induction factor outperforms a simple steady induction decrease in terms of mean power production.
The sharp downward peaks in induction signal appear to stimulate breakdown of the wake and mixing with the free-stream flow. 
Note that mixing here does not refer to turbulent mixing as no turbulence is present in the FVW model.

The use of purely static induction control was previously shown not to be a very effective solution for improving wind farm power production \citep{Annoni2016,VanderHoek2019}.
These results are supported as, within the 2D FVW, the use of under-induction on its own is less effective than the dynamic induction signal acquired through optimisation.
Further study will need to find out whether the combination of slight under-induction and periodic excitation is effective in a realistic wind farm scenario.

The periodic aspect of this control signal resembles the sinusoidal thrust signals that \citet{Munters2018a} found to improve wake mixing by stimulating shedding of vortex rings from the wind turbine.
They find sinusoidal actuation at a non-dimensionalised frequency of $f=0.25$ with a mean local thrust coefficient $c_\mathrm{t}'=2.0$ and amplitude $A=1.5$ to be optimal for turbines operating in a small farm at \SI{5}{D} spacing.
The signal found in the current work has a slightly lower frequency at $f=0.20$, and the induction signal corresponds to a lower mean thrust coefficient $c_\mathrm{t}'=1.75$ with amplitude $A=0.87$.
Their optimisation in a 3D LES environment with turbulent inflow is considerably more complex and more expensive than the 2D FVW, which runs well on a regular laptop.
It is interesting to note that both studies consider a non-rotating actuator-disc wind turbine model.
The differences between the two signals are worth exploring further and will be investigated in future work.


\subsection{Yaw control with wind direction variation}\label{subsec:yaw_results}
Optimisation for yaw control requires the 3D FVW model because it captures the dynamics of the curled wake and therefore shows demonstrable power gain from wake steering, as shown in Figure~\ref{fig:yawsweep}.
The set-up for the optimisation case is as illustrated in Figure~\ref{fig:FVW3D} with the parameters listed in Table~\ref{tab:par_2d}.
A smooth wind direction change with unit magnitude from \SI{0}{\degree} to \SI{-20}{\degree} is implemented to test the capabilities for yaw control under time-varying conditions.
The downstream turbine is assumed to be performing at its greedy optimum with an induction factor $a=0.33$ and a yaw angle that perfectly tracks the inflow direction.
The yaw control signal of the upstream turbine is found as the solution of the optimisation problem.

The objective function \eqref{eq:objective} is constructed with the control signal $\V{m}_k=\begin{bmatrix}
	\psi_k
\end{bmatrix}$, over a prediction horizon of $N_\mathrm{h}=60$ samples.
The output contains the power of both turbines, $\V{y}_k=\begin{bmatrix}
	p_0 & p_1
\end{bmatrix}^\mathrm{T}$.
The objective function weights are set to {$\M{Q}=\begin{bmatrix} -1 & -1 \end{bmatrix}$ and $\M{R}=\begin{bmatrix} 0.025 \end{bmatrix}$}.
This choice of input weight resulted in an adequate balance between input action and power production in an exploratory parameter sweep.
It differs from the 2D case because the yaw control signal has a different magnitude than the induction signal.
At every time step, the optimisation is run for {10} iterations,
after which the first value of the control signal is implemented in the receding horizon control scheme.
Given the slower variations in yaw angle compared to induction control, fewer iterations were required before further iterations no longer yielded consistent improvement in objective function with the current optimiser.
The starting state for the yaw optimisation case is the result of a steady simulation with a \SI{30}{\degree} yaw misalignment on the upstream turbine to reach steady conditions with wake redirection.
This initial condition is illustrated in the first frame of Figure~\ref{fig:vw3d}.

\begin{figure}[!b]
\centering
\includegraphics[width=1.3\linewidth,center]{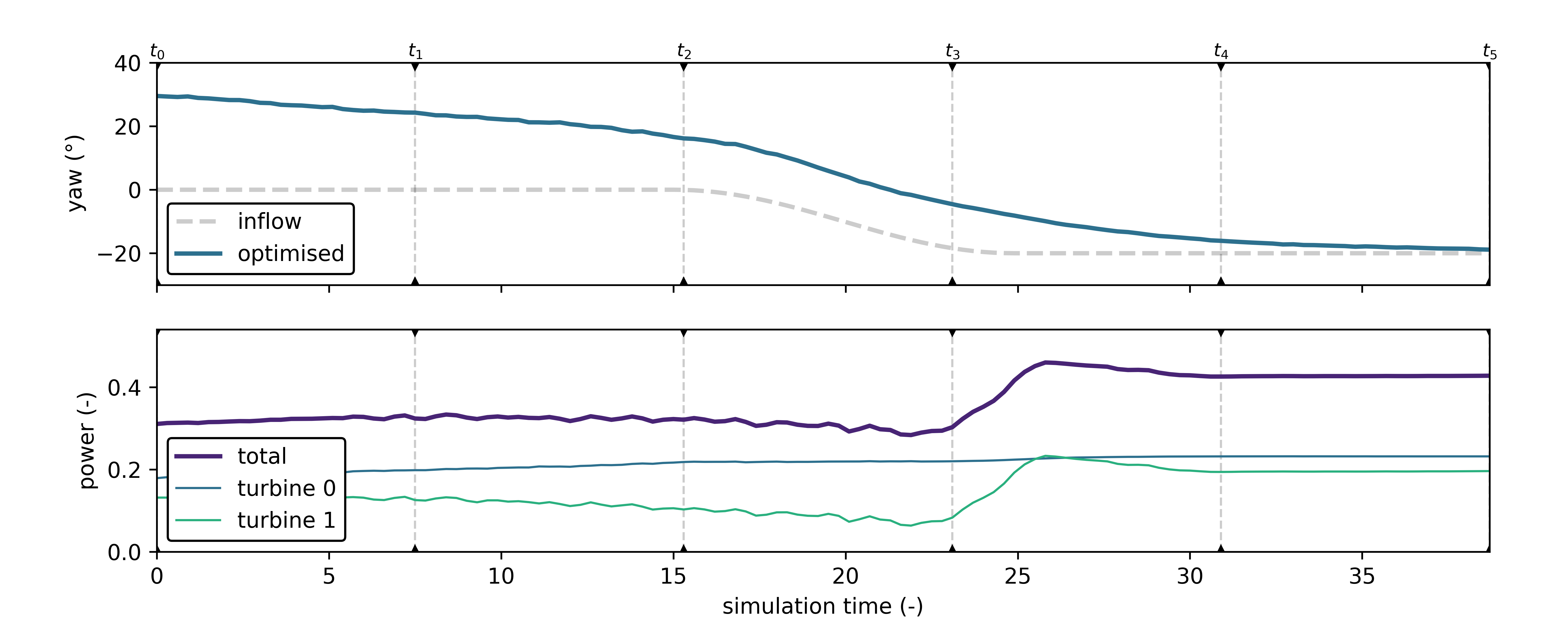}
\caption{
Optimisation of the yaw control signal for total power finds a solution that utilises yaw misalignment to redirect the wake away 
from the downstream turbine.
As the wind direction changes, the turbine is aligned with the flow to perform at its greedy optimum as the wake no longer impinges on the downstream turbine.
Control signal for turbine 0 is optimised whilst turbine 1 is virtually modelled to be operating at its greedy optimum, positioned \SI{5}{D} downstream in fully waked conditions.
Total power is the sum of power from turbine 0 and turbine 1.
Snapshots of the flow field at times $t_0$ to $t_5$ are illustrated in Figure~\ref{fig:vw3d}.
}
\label{fig:wake_steering}
\end{figure}

\begin{figure}[!hbt]
\centering
\includegraphics[width=1.3\linewidth,center]{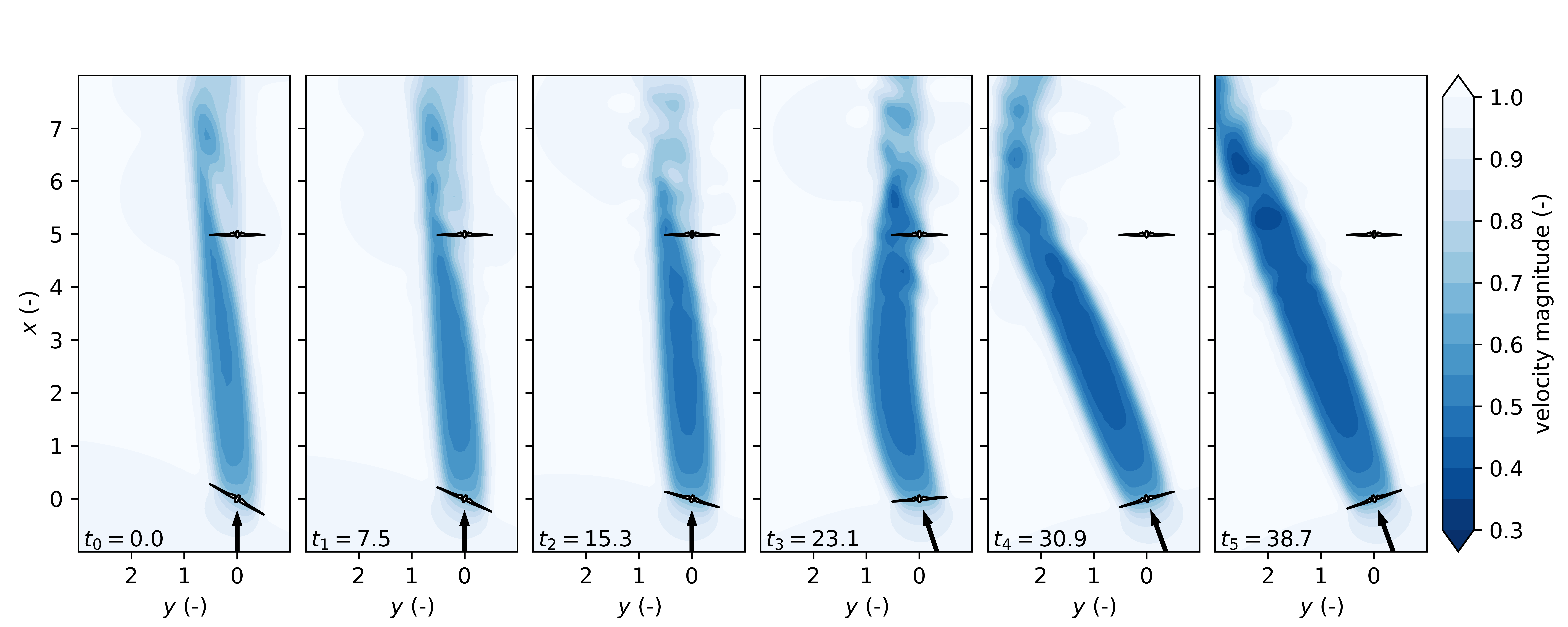}
\caption{
A series of snapshots of the flow field averaged over rotor height for 3D free-vortex wake simulation under rotating inflow.
The black arrow indicates the wind direction and
the yaw signal applied to the turbine is shown in Figure~\ref{fig:wake_steering}.
This case illustrates a transition from wake steering to greedy control as the optimal operating point changes with inflow variation.
}
\label{fig:vw3d} 
\end{figure}

The control signal implemented in the receding horizon control is shown in Figure~\ref{fig:wake_steering} together with the inflow and the associated power production for both turbines.
The result shows wake redirection through yaw misalignment is maintained for the first section where the wind direction has not yet changed.
The turbine is slowly aligned with the inflow in anticipation of the wind direction change.
As the wind direction changes, the upstream turbine is rotated into the wind until it is aligned with the free-stream inflow direction.
The wake no longer interacts with the downstream turbine which makes its greedy optimum a good control solution.

A series of snapshots of the flow field averaged over rotor height are shown in Figure~\ref{fig:vw3d}.
The snapshots illustrate how the change in wind direction propagates through the wake.
It is visible that the turbine control transitions from wake steering with yaw misalignment to greedy control and alignment with the new wind direction.

The use of the 3D FVW as a novel dynamic surrogate model for online control optimisation contrasts with previous work under time-varying wind directions, where
\citet{Campagnolo2020} applied pre-optimised set-points in wind tunnel experiments and
\citet{Doekemeijer2020} used FLORIS to generate steady-state optimal yaw set-points in an online closed-loop controller.
\citet{Howland2020} presented dynamic yaw control using another, more simplified, physics-based model -- a lifting line model with a Gaussian wake.
They operated under unsteady inflow, but with an invariant mean wind direction.
Similarly, unsteady flow without direction changes was considered in the model-free yaw control work by \citet{Ciri2017}. 

Especially for yaw control, the initial guess is critical for attaining good results with the optimisation algorithm.
The optimiser struggles to find good solutions with an initial guess at zero misalignment from a configuration with greedy yaw control and two turbines with full wake interaction.
This seems to be the result of a rather flat optimisation landscape in that configuration.
Given some misalignment in the initial guess for the control signal, the optimiser will tend to find a wake steering solution.
However, when the turbine is currently misaligned to one side, while the other is more effective, the controller is unlikely to switch because the gradient-based optimisation does not cover a large enough search space.
A multi-start optimisation may be a solution to avoid having to predetermine which side to initialise.

Further work is required to validate the effectiveness of the dynamic yaw control results under realistic conditions and 
investigate whether these solutions improve upon wind farm control strategies with steady surrogate models. 
Additionally, a combination of yaw-based control and over-induction has been shown to improve wake steering results~\citep{Cossu2021}, which could be further explored in optimisation with the 3D FVW.

\subsection{
Finite horizon effects in economic MPC
}\label{subsec:empc_discussion}
\begin{figure}[!t]
		\centering
		\includegraphics[width=\linewidth]{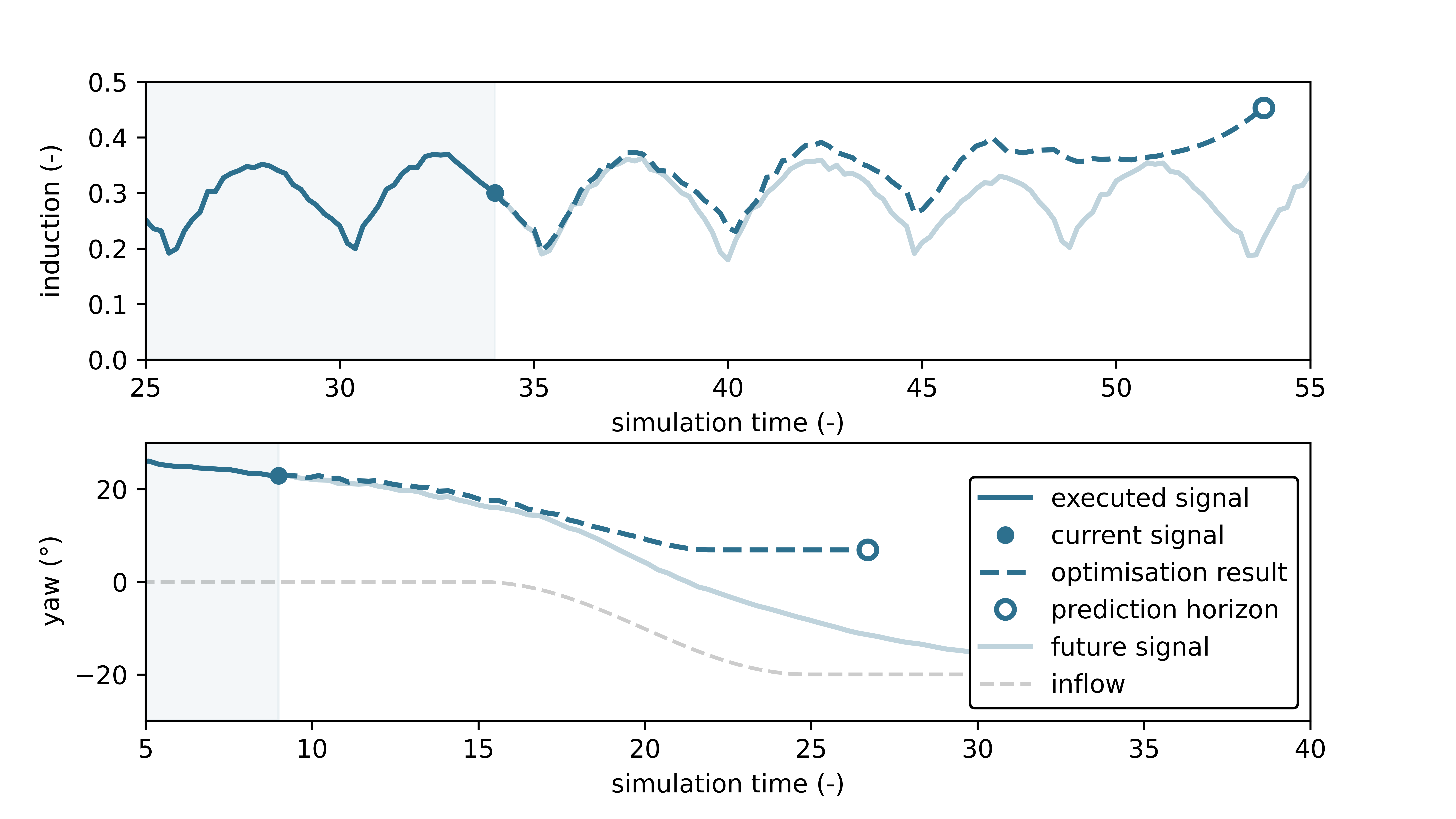}
	\caption{
		The control signal for the finite horizon is found through optimisation at every time-step. The executed signal is the past control input leading up to the current state.
		An optimisation result is found up to the prediction horizon.
		The first step of the optimised signal is implemented, after which it is shifted and re-used at the next time-step to start re-optimisation.
		The future signal is the actual control input that ends up being applied in this receding horizon approach.}
	\label{fig:mpc}
\end{figure}

In addition to the EMPC results in Sections~\ref{subsec:induction_results} and~\ref{subsec:yaw_results}, we illustrate the intermediate results that show the effects of optimisation on a finite horizon.
Figure~\ref{fig:mpc} illustrates the optimisation result for a single optimisation window for both the induction control and the yaw control case.
It illustrates the executed signal, the result of optimisation with a finite horizon at the current time step, and the actual future control signal that is applied in the receding horizon setting.
Towards the finite horizon, the induction returns towards the greedy optimum as the effects no longer reach the downstream turbine.
Right at the horizon, a final peak in induction occurs as the optimiser tries to generate a little more power with over-inductive behaviour.
The effect of optimisation on a finite horizon for yaw control means that the turbine will rotate back to alignment with the inflow when the wake effects no longer propagate to the downstream turbine within the optimisation window.

For the cases presented in this paper, the horizon length is long enough that these end-of-window effects do not affect the control solution as
the receding horizon control scheme prevents these finite horizon effects from being implemented.
However, especially in optimisation cases where the window is relatively short and input cost is high, the horizon effect needs to be properly treated.

%

Horizon length for single wind turbine optimisation should be long enough to push turnpike effects away from the signal to be executed.
In a multiple-turbine setting, horizon length should be chosen at least long enough such that the effects of control signal variation are observed at the downstream turbine for long enough such that the optimiser finds a balanced control solution.
Shorter horizons converge to greedy solutions because  the effects do not propagate to downstream turbines within the optimisation window.



\section{Conclusion}\label{sec:conclusions}
This work presents a control-oriented free-vortex wake (FVW) model of an actuator disc in 2D and 3D to represent a wind turbine wake.
A main novelty in our work is the derivation of the discrete adjoint equations associated with the FVW model, which allows efficient gradient evaluation for use in gradient-based optimisation methods.
The FVW model is computationally efficient enough that the experiments in this paper could be run on a regular laptop computer, without requiring high performance computing clusters.
The evaluation of the gradient requires only a single forward simulation and backward integration of the adjoint states, which is on the order of ten times slower than a simple simulation in the current implementation due to the calculation and storage of all partial derivatives.

The FVW model is implemented as a novel surrogate model for gradient-based control optimisation in an economic model-predictive control setting for maximising mean power production by reducing the negative effects of wind turbine wake interaction.
This implementation allows generation of optimal control solutions and exploration of dynamical wake behaviour.

In a 2D simulation with receding horizon optimisation, dynamic induction control signals are found that combine slight under-induction with a roughly periodic excitation at a frequency of $f=0.20$.
This results in a {\SI{6.0}{\%}} gain over the maximum power generated with steady induction control.
The FVW provides a new, efficient model for exploring dynamic induction control, which has previously been studied in comparatively expensive LES studies.

The 3D free-vortex wake model exhibits curled wake dynamics under yaw misalignment and is therefore suitable as a novel physics-based surrogate model for dynamic wake steering control.
The economic model-predictive control strategy finds yaw signals under time-varying wind direction that show both wake redirection and a smooth return to greedy control.
It contrasts existing literature that bases yaw control on steady model results or considers dynamic control with unsteady flow, but invariant mean wind directions.

Future work will include validation of the model response, exploration of the dynamics of the optimisation problem and further experiments with the optimisation of induction control and wake redirection.


\appendix
\section{Partial Derivatives}\label{sec:derivatives}
This section presents the partial derivatives required for the calculation of the gradient with the discrete adjoint.
Derivatives of the model state update from Section~\ref{sec:model} are provided in \ref{subsec:derivative_state} and those of the output and objective function from Section~\ref{sec:optimisation} are provided in \ref{subsec:derivative_objective}.

\subsection{State update}\label{subsec:derivative_state}
The Jacobian matrix of the state update with respect to the previous state  can be written as
\begin{align}
\frac{\partial{f}(\V{q}_k,\V{m}_k)}{\partial \V{q}_{k}} =
\frac{\partial\V{q}_{k+1}}{\partial \V{q}_{k}} = 
\frac{\partial}{\partial \V{q}_{k}} 
\begin{bmatrix}
	\V{X} \\ \V{\Gamma} \\ \V{U} \\ \V{M}
\end{bmatrix}_{k+1} 
	&= \begin{bmatrix}
		\frac{\partial \V{X}_{k+1}}{\partial \V{X}_{k}}
			& \frac{\partial \V{X}_{k+1}}{\partial \V{\Gamma}_k}
			& \frac{\partial \V{X}_{k+1}}{\partial \V{U}_k}
			&	\M{0} \\
		\frac{\partial \V{\Gamma}_{k+1}}{\partial \V{X}_{k}}
			& \frac{\partial \V{\Gamma}_{k+1}}{\partial \V{\Gamma}_k}
			&\frac{\partial \V{\Gamma}_{k+1}}{\partial \V{U}_k}
			&	\M{0}\\ 
			\M{0}
			& \M{0} 
			& \frac{\partial \V{U}_{k+1}}{\partial \V{U}_k}
			& \M{0}\\
			\M{0}
			& \M{0} 
			& \M{0} 
			&	\M{0}\\
	\end{bmatrix}\,,
\end{align}
where those derivatives that are zero have been removed, leaving a number of sub-matrices to be constructed.

First, take the partial derivatives of the induced velocity  of a single vortex element $\V{u}_\mathrm{i}$ in 2D,
which is divided in three parts ($\V{u}_0$, $u_1$ and $u_2$) to simplify calculation,
\begin{align}
\V{u}_\mathrm{i}(\V{x}_0, \V{x}_1) = 
\underbrace{\begin{bmatrix}-r_y\\r_x\end{bmatrix}}_{\V{u}_0}
\underbrace{\left(
\frac{\Gamma}{2\pi}
\frac{1}{||\V{r}||^2}
\right)}_{{u}_1}
\underbrace{
\left(
1 - \exp\left(-\frac{||\V{r}||^2}{\sigma^2}\right)
\right)}_{{u}_2} \,.
\end{align}
For clarity, this section will refer to $\V{u}_\mathrm{i}$ as $\V{u}$.

The derivative of induced velocity with respect to positions is
\begin{align}
	\frac{\partial\V{u}}{\partial \V{x}_0} = -\frac{\partial\V{u}}{\partial \V{r}} \,, \quad
	\frac{\partial\V{u}}{\partial \V{x}_1} = \frac{\partial\V{u}}{\partial \V{r}}\,,
\end{align}
where the derivative to the relative position can be expanded as
\begin{align}
	\frac{\partial{\V{u}}}{\partial \V{r}} &=
	\frac{\partial\V{u}_0}{\partial \V{r}}{u}_1 {u}_2
	+ \V{u}_0\frac{\partial{u}_1}{\partial \V{r}}  {u}_2
	+ \V{u}_0 {u}_1 \frac{\partial{u}_2}{\partial \V{r}}\,.
\end{align}
The required partial derivatives are
\begin{align}
\frac{\partial\V{u}_0}{\partial \V{r}} &= \begin{bmatrix}
	0 & -1 \\ 1 & 0\\
\end{bmatrix}\,,\\
 \frac{\partial {u}_1}{\partial \V{r}} &= \frac{\Gamma}{\pi}
		\left(\frac{\V{r}^{\mathrm{T}}}
		{||\V{r}||^4}		\right) \,,\\
\frac{\partial {u}_2}{\partial \V{r}} &= 
\frac{2 \V{r}^{\mathrm{T}}}{\sigma^2} 
\exp\left(-\frac{||\V{r}||^2}{\sigma^2}\right) \,.
\end{align}
For all rings $b>0$, the Jacobian matrix from element positions to positions is filled with the partial derivatives as 
\begin{align}
\begin{bmatrix}
	\frac{\partial\V{X}^{(b)}}{\partial \V{X}^{(a)}}
\end{bmatrix}_{ij} &= 
			h\frac{\partial \V{u}_{i}^{(b)}}{\partial \V{x}_j^{(a)}}
			\left(\V{x}_i^{(b)}, \V{x}_{j}^{(a)}\right)
\\
	&\text{for } i,j=0,1 \text{ , } a=1,2,\dots,n_\mathrm{r} \text{ and } b=1,2,\dots,n_\mathrm{r}-1\,. \nonumber
\end{align}
If $a+1=b$, then 
the diagonal elements are substituted as
\begin{align}
\begin{bmatrix}
	\frac{\partial\V{X}_{k+1}^{(b)}}{\partial \V{X}_{k}^{(b-1)}}
\end{bmatrix}_{ii} \leftarrow  
	1 
	+ h\sum_{m=1}^{n_{\mathrm{r}-1}} \sum_{n=0}^{1}
	\frac{\partial \V{u}_{i}^{(b-1)}}{\partial \V{x}_j^{(b-1)}}
		\left(\V{x}_i^{(b-1)}, \V{x}_{n}^{(m)}\right)\,.  
\end{align}

The induced velocity $\V{u}_\mathrm{i}$ in 3D is similarly divided into three parts $\V{u}_0$, ${u}_1$, and ${u}_2$,
\begin{align}
\V{u}_\mathrm{i}(\V{x}_0, \V{x}_1, \V{x}_2) = 
\underbrace{\left(
\frac{\Gamma}{4\pi}
\frac{\V{r}_{1}\times \V{r}_{2}}{||\V{r}_{1}\times\V{r}_{2}||^2}
\right)}_{\V{u}_0}
\underbrace{
\left(
\V{r}_{0}\cdot
\left(
\frac{\V{r}_{1}}{||\V{r}_{1}||}
- \frac{\V{r}_{2}}{||\V{r}_{2}||}
\right)
\right)}_{u_1}
\underbrace{
\left(
1 - \exp\left(-\frac{||\V{r}_1 \times \V{r_2}||^2}{\sigma^2||\V{r}_0||^2}\right)
\right)}_{u_2}\,.
\end{align}
%

%
The required full derivatives of induced velocity with respect to the position inputs can be expanded as
\begin{align}
	\frac{\partial{\V{u}}}{\partial \V{x}_0} &= 
		- \frac{\partial{\V{u}}}{\partial \V{r}_1} 
		- \frac{\partial{\V{u}}}{\partial \V{r}_2} \,,\\
	\frac{\partial{\V{u}}}{\partial \V{x}_1} &= 
		\frac{\partial{\V{u}}}{\partial \V{r}_1}
		- \frac{\partial{\V{u}}}{\partial \V{r}_0} \,,\\
	\frac{\partial{\V{u}}}{\partial \V{x}_2} &= 
		\frac{\partial{\V{u}}}{\partial \V{r}_2} 
		+\frac{\partial{\V{u}}}{\partial \V{r}_0}  \,.
\end{align}
We expand the required derivatives according to the product rule into manageable parts,
\begin{align}
	\frac{\partial{\V{u}}}{\partial \V{r}_0} &=
	\frac{\partial\V{u}_0}{\partial \V{r}_0}{u}_1 {u}_2
	+ \V{u}_0\frac{\partial{u}_1}{\partial \V{r}_0}  {u}_2
	+ \V{u}_0 {u}_1 \frac{\partial{u}_2}{\partial \V{r}_0}\,,\\
%
	\frac{\partial{\V{u}}}{\partial \V{r}_1}&=
	\frac{\partial\V{u}_0}{\partial \V{r}_1} {u}_1 {u}_2
	+ \V{u}_0\frac{\partial{u}_1}{\partial \V{r}_1}  {u}_2
	+ \V{u}_0 {u}_1 \frac{\partial{u}_2}{\partial \V{r}_1}\,,\\
%
	\frac{\partial{\V{u}}}{\partial \V{r}_2}	&=
	\frac{\partial\V{u}_0}{\partial \V{r}_2} {u}_1 {u}_2
	+ \V{u}_0\frac{\partial {u}_1}{\partial {r}_2}  {u}_2
	+ \V{u}_0 {u}_1 \frac{\partial{u}_2}{\partial \V{r}_2}\,.
\end{align}
%
The partial derivatives of $\V{u}_0$ are
\begin{align}
		\frac{\partial\V{u}_0}{\partial \V{r}_0} 
		&= \M{0} \,,\\
		\frac{\partial\V{u}_0}{\partial \V{r}_1} 	
		&=  \frac{\Gamma}{4\pi}
		\left(\frac{||\V{r}_{1}\times\V{r}_{2}||^2 \M{I}_3 - 2\left(\V{r}_{1}\times \V{r}_{2}\right)\left(\V{r}_{1}\times \V{r}_{2}\right)^{\mathrm{T}}}
		{||\V{r}_{1}\times\V{r}_{2}||^4}		\right)
		\left[\V{r}_2	\right]_{\times}^{\mathrm{T}} \,,\\
		\frac{\partial\V{u}_0}{\partial \V{r}_2}& = 
		 \frac{\Gamma}{4\pi}
		\left(\frac{||\V{r}_{1}\times\V{r}_{2}||^2\M{I}_3 - 2\left(\V{r}_{1}\times \V{r}_{2}\right)\left(\V{r}_{1}\times \V{r}_{2}\right)^{\mathrm{T}}}
		{||\V{r}_{1}\times\V{r}_{2}||^4}		\right)
		\left[\V{r}_1	\right]_{\times} \,.
\end{align}
	Here, $[\V{a}]_{\times}$ indicates the skew-symmetric constructed from a vector $\V{a}$, such that the cross product can be written in the form
	\begin{align}
		\V{a} \times \V{b} = [\V{a}]_{\times}\V{b} = \begin{bmatrix}
0 & -a_3 & a_2 \\
a_3 & 0 & -a_1 \\
-a_2 & a_1 & 0 \\
\end{bmatrix}\begin{bmatrix}
b_1\\b_2 \\ b_3\\
\end{bmatrix}\,,
	\end{align}
	allowing compact notation of the derivative
	\begin{align}
		\frac{\partial\V{a}\times \V{b}}{\partial\V{b}} = [\V{a}]_{\times}\,.
	\end{align}
The partial derivatives of $u_1$ are
\begin{align}
		\frac{\partial{u}_1}{\partial \V{r}_0} 
		& =	\left(
						\frac{\V{r}_{1}}{||\V{r}_{1}||}
						- \frac{\V{r}_{2}}{||\V{r}_{2}||}
						\right)^{\mathrm{T}} \,, \\
		\frac{\partial{u}_1}{\partial \V{r}_1} & = \frac{||\V{r}_1||^2\V{r}_0^\mathrm{T} - (\V{r}_0\cdot\V{r}_1)\V{r}_1^\mathrm{T}}{||\V{r}_1||^3} \,, \\
		\frac{\partial{u}_1}{\partial \V{r}_2} & = -\frac{||\V{r}_2||^2\V{r}_0^\mathrm{T} - (\V{r}_0\cdot\V{r}_2)\V{r}_2^\mathrm{T}}{||\V{r}_2||^3}\,.
\end{align}
Finally, the partial derivatives of $u_2$ are
\begin{align}
		\frac{\partial{u}_2}{\partial \V{r}_0} 
			& = -\exp\left(-\frac{||\V{r}_1 \times \V{r_2}||^2}{\sigma^2||\V{r}_0||^2}\right)
		\left(2\frac{||\V{r}_1 \times \V{r_2}||^2}{\sigma^2 ||\V{r}_0||^4}\right) {\V{r}_0^{\mathrm{T}}}\,,\\
		\frac{\partial{u}_2}{\partial \V{r}_1} 
	& = \exp\left(-\frac{||\V{r}_1 \times \V{r_2}||^2}{\sigma^2||\V{r}_0||^2}\right)
		\left(
		\frac{2\left(\V{r}_1 \times \V{r_2}\right)^{\mathrm{T}}}{\sigma^2||\V{r}_0||^2}
		\right)
		\left[\V{r}_2\right]_{\times}^{\mathrm{T}} \,,\\
		\frac{\partial{u}_2}{\partial \V{r}_2}
		& = \exp\left(-\frac{||\V{r}_1 \times \V{r_2}||^2}{\sigma^2||\V{r}_0||^2}\right)
		\left(
		\frac{2\left(\V{r}_1 \times \V{r_2}\right)^{\mathrm{T}}}{\sigma^2||\V{r}_0||^2}
		\right)
		\left[\V{r}_1\right]_{\times} \,.
\end{align}
For all rings $b>0$, the Jacobian matrix is filled with the partial derivatives as 
\begin{align}
\begin{bmatrix}
	\frac{\partial\V{X}^{(b)}}{\partial \V{X}^{(a)}}
\end{bmatrix}_{ij} &= \left\{
	\begin{array}{ll}
			h\frac{\partial \V{u}_{i}^{(b)}}{\partial \V{x}_j^{(a)}}
			\left(\V{x}_i^{(b)}, \V{x}_{j}^{(a)}, \V{x}_{j+1}^{(a)}\right)
			& \text{if } j=0\,,\\
		h\frac{\partial \V{u}_{i}^{(b)}}{\partial \V{x}_j^{(a)}}
		\left(\V{x}_i^{(b)}, \V{x}_{j-1}^{(a)}, \V{x}_{j}^{(a)}\right)
			& \text{if } j=n_\mathrm{e}\,,\\
		 h\frac{\partial \V{u}_{i}^{(b)}}{\partial \V{x}_j^{(a)}}
		\left(\V{x}_i^{(b)}, \V{x}_{j-1}^{(a)}, \V{x}_{j}^{(a)}\right)
			+h\frac{\partial \V{u}_{i}^{(b)}}{\partial \V{x}_j^{(a)}}
			\left(\V{x}_i^{(b)}, \V{x}_{j}^{(a)}, \V{x}_{j+1}^{(a)}\right)
				& \text{otherwise}\,,
	\end{array}	
	\right.	\\
	&\text{for } i,j=0,1,\dots,n_\mathrm{e} \text{ , } a=0,1,\dots,n_\mathrm{r}-1 \text{ and } b=1,2,\dots,n_\mathrm{r}-1\,. \nonumber
\end{align}
If $a+1=b$, then 
the diagonal elements are substituted as
\begin{align}
\begin{bmatrix}
	\frac{\partial\V{X}_{k+1}^{(b)}}{\partial \V{X}_{k}^{(b-1)}}
\end{bmatrix}_{ii} \leftarrow  
	1 
	+ h\sum_{m=1}^{n_{\mathrm{r}-1}} \sum_{n=1}^{n_{\mathrm{e}}}
	\frac{\partial \V{u}_{i}^{(b-1)}}{\partial \V{x}_j^{(b-1)}}
		\left(\V{x}_i^{(b-1)}, \V{x}_{n-1}^{(m)}, \V{x}_{n}^{(m)}\right)\,.  
\end{align}

The position of the first ring depends only on the yaw angle, therefore
\begin{align}
\frac{\partial\V{X}^{(0)}}{\partial \V{X}^{(a)}} = \M{0}\,.
\end{align}
%

Further derivatives are mostly independent from the dimension of the FVW model used.

The position update for all rings $b>0$ depends on the vortex strength of all elements as
\begin{align}
	\begin{bmatrix}
		\frac{\partial \V{X}_{k+1}^{(b)}}{\partial \V{\Gamma}_k^{(a)}} 
	\end{bmatrix}_{ij}&= 
		\frac{\partial\V{x}_i^{(b)}}{\partial \V{u}_i^{(b-1)}}\frac{\partial\V{u}_i^{(b-1)}}{\partial \Gamma_n^{(a)}} =
		h\frac{\partial\V{u}_i^{(b-1)}}{\partial \Gamma_j^{(a)}} 
		\left(\V{x}_i^{(b-1)}, \V{x}_{j-1}^{(a)}, \V{x}_{j}^{(a)}\right)
		\,\\
		&\text{for } i=0,1,\dots,n_\mathrm{e} \text{ , } j=1,2,\dots,n_\mathrm{e} \text{ and } a = 0,1,\dots,n_\mathrm{r}-1 \,, \nonumber
\end{align}
where the partial derivative of induced velocity to vortex strength is
\begin{align}
\frac{\partial{\V{u}_i}}{\partial \Gamma_j} 
= \frac{1}{\Gamma}_j \V{u}_i\,.
\end{align}
The points defining the vortex filament travel based on their stored free-stream velocity, so for all rings $b>0$,
\begin{align}
	{
	\frac{\partial \V{X}^{(b)}_{k+1}}{\partial \V{U}_k^{(b-1)}}
	}
	 = h \M{I}_{n_\mathrm{p}} \,.
\end{align}
Vortex strength and free-stream velocity are passed from one ring to the next for all rings $b>0$,
%
\begin{align}
\frac{\partial \V{\Gamma}_{k+1}^{(b)}}{\partial \V{\Gamma}_k^{(b-1)}} &= \M{I}_{n_\mathrm{e}} \,,\\
	\frac{\partial \V{U}_{k+1}^{(b)}}{\partial \V{U}_{k}^{(b-1)}} &= \M{I}_{n_\mathrm{p}} \,.
\end{align}

The initialisation of a new ring depends on the entire state of the model through the rotor velocity,
\begin{align}
	{
	\frac{\partial \V{\Gamma}_{k+1}^{(0)}}{\partial \V{X}_k}
	}
	 &= \frac{\partial \V{\Gamma}^{(0)}}{\partial \V{u}_\mathrm{r}}
	 \frac{\partial \V{u}_\mathrm{r}}{\partial \V{X}}\,,\\
	 {
	\frac{\partial \V{\Gamma}^{(0)}_{k+1}}{\partial \V{\Gamma}_k}
	} 
	 &= \frac{\partial \V{\Gamma}^{(0)}}{\partial \V{u}_\mathrm{r}}
	 \frac{\partial \V{u}_\mathrm{r}}{\partial \V{\Gamma}} \,,\\
	{
	\frac{\partial \V{\Gamma}^{(0)}_{k+1}}{\partial \V{U}_k}
	}
	 &= \frac{\partial \V{\Gamma}^{(0)}}{\partial \V{u}_\mathrm{r}}
	 \frac{\partial \V{u}_\mathrm{r}}{\partial \V{U}}\,,
\end{align}
where
\begin{align}
	\frac{\partial \V{\Gamma}}{\partial \V{u}_\mathrm{r}}
	 =  h c_{t}'(a) \V{n}(\psi) (\V{u}_\mathrm{r}^{\operatorname{T}}\V{n}(\psi))\,.
\end{align}
The Jacobian of rotor speed with respect to all element positions is calculated as
\begin{align}
	\frac{\partial \V{u}_r}{\partial \V{X}_k}
		&= \frac{1}{n_\mathrm{u}}\sum_{i=0}^{n_\mathrm{u}-1} 
			\left(\frac{\partial}{\partial \V{X}_k}
			\left(\V{u}_\mathrm{\infty}(\V{x}_i,\V{q})\right) +\frac{\partial}{\partial \V{X}_k}
			\left(\V{u}_\mathrm{ind}(\V{x}_i,\V{q})\right) \right)\,,
\end{align}
with
\begin{align}
	{
	\frac{\partial \V{u}_\mathrm{\infty}(\V{x}_i,\V{q})}{\partial \V{X}} =
	 20 (\V{x}-\V{x}_i^{(b)}) \bar{w}_i^{(b)} -
	 \bar{w}_i^{(b)}\sum_{i=0}^{n_\mathrm{u}-1} \sum_{b=0}^{n_\mathrm{r}}  20(\V{x}-\V{x}_i^{(b)})\bar{w}_i^{(b)}}\,,
\end{align}
with the partial derivatives placed in 2D as
\begin{align}
\begin{bmatrix}
	\frac{\partial\V{u}_\mathrm{ind}(\V{x}_i,\V{q})}{\partial \V{X}^{(a)}}
\end{bmatrix}_{0j} &= 
		\frac{\partial \V{u}}{\partial \V{x}_j^{(a)}}
			\left(\V{x}_i, \V{x}_{j}^{(a)}\right)
\\
	&\text{for } j=0,1 \text{ , } a=0,1,\dots,n_\mathrm{r}-1\,,\nonumber
\end{align}
and for the 3D FVW as
\begin{align}
\begin{bmatrix}
	\frac{\partial\V{u}_\mathrm{ind}(\V{x}_i,\V{q})}{\partial \V{X}^{(a)}}
\end{bmatrix}_{0j} &= \left\{
	\begin{array}{ll}
		\frac{\partial \V{u}}{\partial \V{x}_j^{(a)}}
			\left(\V{x}_i, \V{x}_{j}^{(a)}, \V{x}_{j+1}^{(a)}\right)
			& \text{if } j=0\,,\\
		\frac{\partial \V{u}}{\partial \V{x}_j^{(a)}}
		\left(\V{x}_i, \V{x}_{j-1}^{(a)}, \V{x}_{j}^{(a)}\right)
			& \text{if } j=n_\mathrm{e}\,,\\
		 \frac{\partial \V{u}}{\partial \V{x}_j^{(a)}}
		\left(\V{x}_i, \V{x}_{j-1}^{(a)}, \V{x}_{j}^{(a)}\right)
			+\frac{\partial \V{u}}{\partial \V{x}_j^{(a)}}
			\left(\V{x}_i, \V{x}_{j}^{(a)}, \V{x}_{j+1}^{(a)}\right)
				& \text{otherwise}\,,
	\end{array}	
	\right.	\\
	&\text{for } j=0,1,\dots,n_\mathrm{e} \text{ , } a=0,1,\dots,n_\mathrm{r}-1\,. \nonumber
\end{align}
The Jacobian with respect to all vortex strengths is calculated as
\begin{align}
	\frac{\partial \V{u}_\mathrm{r}}{\partial \V{\Gamma}_k}
		&= \frac{1}{n_\mathrm{u}}\sum_{i=0}^{n_\mathrm{u}-1} 
			\frac{\partial}{\partial \V{\Gamma}_k}
			\left(\V{u}_\mathrm{ind}(\V{x}_i,\V{q})\right) \,,
\end{align}
with the partial derivatives
\begin{align}
	\begin{bmatrix}
		\frac{\partial \V{u}_\mathrm{ind}(\V{x}_i,\V{q})}{\partial \V{\Gamma}_k^{(a)}} 
	\end{bmatrix}_{1j}&= 
		\frac{\partial\V{u}}{\partial \Gamma_j^{(a)}} 
		\left(\V{x}_i, \V{x}_{j-1}^{(a)}, \V{x}_{j}^{(a)}\right)
		\,\\
		&\text{for }  j=1,2,\dots,n_\mathrm{e} \text{ and } a = 0,1,\dots,n_\mathrm{r}-1 \,. \nonumber
\end{align}
The derivative of the free-stream contribution to the disc-averaged velocity over the rotor is
\begin{align}
	{
	\frac{\partial \V{u}_\mathrm{r}}{\partial \V{u}_{\infty,i}^{(b)}}
	} =  \frac{1}{n_\mathrm{u}}\sum_{i=0}^{n_\mathrm{u}-1}    \bar{w}_i^{(b)}(\V{x}_i,\V{q})\,.
\end{align}

The Jacobian matrix of the state update with respect to the inputs is defined as
\begin{align}
\frac{\partial\V{q}_{k+1}}{\partial \V{m}_{k}} = 
\frac{\partial}{\partial \V{m}_{k}} 
\begin{bmatrix}
	\V{X} \\ \V{\Gamma} \\ \V{U} \\ \V{M}
\end{bmatrix}_{k+1} =
 \begin{bmatrix}
		\frac{\partial \V{X}_{k+1}}{\partial \V{m}_{k}}\\
		\frac{\partial \V{\Gamma}_{k+1}}{\partial \V{m}_{k}}\\
			\M{0}\\
			\frac{\partial \V{M}_{k+1}}{\partial \V{m}_{k}}
	\end{bmatrix}\,.
\end{align}

The position of the first ring is directly controlled with the yaw angle
\begin{align}
	\begin{bmatrix}
	 	\frac{\partial \V{X}^{(0)}}{\partial \psi}
	 \end{bmatrix}_i = \frac{\partial\V{x}_i^{(0)}}{\partial\psi}
 \quad \text{for } i={0,1,\dots,n_\mathrm{e}} \,,
\end{align}
which in 2D yields
\begin{align}
\frac{\partial\V{x}_0^{(0)}}{\partial\psi}
 = \frac{\partial\M{R}_z(\psi_k)}{\partial \psi}\begin{bmatrix}
0 \\ r
\end{bmatrix}\,,\quad
\frac{\partial\V{x}_1^{(0)}}{\partial\psi}
 = \frac{\partial\M{R}_z(\psi_k)}{\partial \psi}(\psi_k)\begin{bmatrix}
0 \\ -r
\end{bmatrix}\,,
\end{align}
and in 3D yields
\begin{align}
\frac{\partial \V{x}_i^{(0)}(\psi)}{\partial \psi}
 = \frac{\partial\M{R}_z(\psi_k)}{\partial \psi}\begin{bmatrix}
0 \\ r\cos(2\pi \frac{i}{n_\mathrm{e}}) \\ r\sin(2\pi \frac{i}{n_\mathrm{e}})
\end{bmatrix} \quad \text{for } i={0,1,\dots,n_\mathrm{e}} \,,
\end{align}
given the derivative of the rotation matrix in the relevant dimension,
\begin{align}
		\frac{\partial \M{R}_z(\psi)}{\partial\psi}=
	\begin{bmatrix}
		-\sin\psi & \cos\psi\\
		-\cos \psi & -\sin\psi
	\end{bmatrix}\,,\quad\frac{\partial \M{R}_z(\psi)}{\partial\psi}=
	\begin{bmatrix}
		-\sin\psi & \cos\psi & 0 \\
		-\cos \psi & -\sin\psi & 0\\
		0 & 0 & 0\\
	\end{bmatrix}\,.
\end{align}
%



The vortex strength of the first ring depends on the yaw angle as
\begin{align}
	\begin{bmatrix}
		\frac{\partial \V{\Gamma}^{(0)}}{\partial \psi}
	\end{bmatrix}_i = 
\frac{\partial \Gamma_i^{(0)}}{\partial \psi} \left(a, \psi \right)
\quad \text{for } i=1,2,\dots,n_{\mathrm{e}} \,,
\end{align}
where
\begin{align}
	\frac{\partial \Gamma_i^{(0)}}{\partial \psi}
	&= h c_\mathrm{t}'(a) (\V{u}_r^{\operatorname{T}}\V{n}(\psi))\left(
	\V{u}_\mathrm{r}^{\operatorname{T}} \frac{\partial \V{n}(\psi) }{\partial \psi}
	\right)\,.
\end{align}

The vortex strength of the first ring depends on the thrust coefficient as
\begin{align}
	\begin{bmatrix}
		\frac{\partial \V{\Gamma}^{(0)}}{\partial a}
	\end{bmatrix}_i = 
\frac{\partial \Gamma_i^{(0)}}{\partial a} \left(a, \psi\right)
\quad \text{for } i=1,2,\dots,n_{\mathrm{e}} \,,
\end{align}
where
\begin{align}
	\frac{\partial \Gamma_i^{(0)}}{\partial a}
	&=\left(h \frac{1}{2} (\V{u}_{\mathrm{r}}\cdot\V{n}(\psi))^2 \right) \frac{\partial c_\mathrm{t}'}{\partial a} \,,\\
   \frac{\partial c_\mathrm{t}'}{\partial a} &= \left\{\begin{array}{ll}
	\frac{4}{(1-a)^2} 	& \text{if } a\leq a_\mathrm{t}\,,\\
	\frac{2c_\mathrm{t1}}{(1-a)^3} - \frac{4(\sqrt{c_\mathrm{t1}}-1)}{(1-a)^2}
	& \text{if } a>a_\mathrm{t}\,.
	\end{array}
	\right.
\end{align}
The derivative of saved controls is identity,
\begin{align}
	\frac{\partial \V{M}_{k+1}}{\partial \V{m}_k}
	 = \M{I}_{n_\mathrm{c}}\,.
\end{align}

\subsection{Output and objective function}\label{subsec:derivative_objective}
The input sensitivity of the objective function is calculated as
\begin{align}
	\frac{\partial J_k}{\partial \V{q}_k} 
	&= \M{Q}\frac{\partial\V{y}_k}{\partial \V{q}_k} - 2\V{\Delta m}_k^\mathrm{T} \M{R} \,,\\
	\frac{\partial J_k}{\partial \V{m}_k} 
	&= \M{Q}\frac{\partial\V{y}_k}{\partial \V{m}_k}+  2\V{\Delta m}_k^\mathrm{T} \M{R}\,.
\end{align}
The partial derivative of the output to the controls is $\partial \V{y}_k/\partial \V{m}_k=0 $ because the power is calculated with controls saved in the state vector $\V{q}_k$.
The Jacobian of the output to the states is given as
\begin{align}
	\frac{\partial\V{y}_k}{\partial \V{q}_k} 
	= \begin{bmatrix}
		\frac{\partial \V{y}_k}{\partial \V{u}_\mathrm{r}}
	\end{bmatrix}
	\begin{bmatrix}
		\frac{\partial \V{u}_\mathrm{r}}{\partial \V{X}_k}
		 & \frac{\partial \V{u}_\mathrm{r}}{\partial \V{\Gamma}_k}
		 & \frac{\partial \V{u}_\mathrm{r}}{\partial \V{U}_k}
		 & \frac{\partial \V{u}_\mathrm{r}}{\partial \V{M}_k}
		\end{bmatrix} \nonumber
	+ \begin{bmatrix}
		\M{0}
		 & \M{0}
		 & \M{0}
		 & \frac{\partial \V{y}_k}{\partial \V{M}_k}
		\end{bmatrix}\,.
\end{align}

The Jacobian of the power output to the saved control inputs for the two-turbine case presented in this paper is constructed as
\begin{align}
	\frac{\partial\V{y}_k}{\partial \V{M}_k} = \begin{bmatrix}
		\frac{\partial P_0}{\partial a_0} 
			& \frac{\partial P_0}{\partial \psi_0} &0&0 \\
		0 & 0 &\frac{\partial P_1}{\partial a_1}&
			 \frac{\partial P_1}{\partial \psi_1} 
		\end{bmatrix}\,,
\end{align}
with the partial derivatives to the axial induction
\begin{align}
	\frac{\partial P_0}{\partial a_0} 
	&= 
	\frac{1}{2} A_\mathrm{r} (\V{u}_{\mathrm{r}_0}\cdot \V{n}(\psi_0))^3
		\frac{4}{(1-a_0)^2}\,,\\
	{\frac{\partial P_1}{\partial a_1}} &= 
		\frac{1}{2} A_\mathrm{r} (\V{u}_{\mathrm{r}_1}\cdot \V{n}(\psi_1))^3
		(-8a_1 (1-a_1)+4(1-a_1)^2)\,,
\end{align}
and the partial derivatives to the yaw angle
\begin{align}
	\frac{\partial P_0}{\partial \psi_0}
	&=\frac{3}{2}c_\mathrm{P}'(a_0) A_\mathrm{r} (\V{u}_{\mathrm{r}_0}\cdot \V{n}(\psi_0))^2 (\V{u}_{\mathrm{r}_0}^\mathrm{T}
		\frac{\partial \V{n}(\psi_0)}{\partial \psi_0} )\,,\\
	\frac{\partial P_1}{\partial \psi_1}&=\frac{3}{2}c_\mathrm{P}'(a_1) A_\mathrm{r} (\V{u}_{\mathrm{r}_1}^* \cdot \V{n}(\psi_1))^2 (\V{u}_{\mathrm{r}_1}^\mathrm{*T}
		\frac{\partial \V{n}(\psi_1)}{\partial \psi_1} )\,.
\end{align}
%
Taking the derivative of the output to the rotor velocity yields
\begin{align}
	\begin{bmatrix}
		\frac{\partial \V{y}_k}{\partial \V{u}_\mathrm{r}}
	\end{bmatrix}
	& = \begin{bmatrix}
		\frac{\partial P_0}{\partial \V{u}_{\mathrm{r}_0}} & 0\\
		0 & \frac{\partial P_1}{\partial \V{u}_{\mathrm{r}_1}}\\
	\end{bmatrix}\,,\\
	\frac{\partial P_0}{\partial \V{u}_{\mathrm{r}_0}}
	&={\frac{3}{2}c_\mathrm{P}'(a_0) A_\mathrm{r} (\V{u}_{\mathrm{r}_0}\cdot \V{n}(\psi_0))^2 \V{n}(\psi_0)^\mathrm{T}} \,, \\
	\frac{\partial P_1}{\partial \V{u}_{\mathrm{r}_1}}
	&={\frac{3}{2}c_\mathrm{P}'(a_1) A_\mathrm{r} (\V{u}_{\mathrm{r}_1}\cdot \V{n}(\psi_1))^2 \V{n}(\psi_1)^\mathrm{T}(1-a_1)\,.}
\end{align}
%

%

\section{Convergence Study for {Numerical Methods}}\label{sec:parameter_choice}
A variation of model parameters {for the numerical methods} was performed to show convergence for the model configuration 
chosen in this paper as presented in Section~\ref{sub:steady}.
The power output under steady conditions is compared to the reference power from momentum theory for validation, calculated as
\begin{align}
	p_{0,\mathrm{ref}} &= c_{\mathrm{p}}'(a_0) A_{\mathrm{r}} (u_\infty (1-a_0))^3  \,,\\
	p_{1,\mathrm{ref}} &= c_{\mathrm{p}}'(a_1) A_{\mathrm{r}} (u_\infty (1-2a_0) (1-a_1))^3 \,,
\end{align}
and normalised as $\bar{p} = p/p_{\mathrm{ref}}$.
Turbine configuration is chosen with a \SI{5}{D} spacing as used in this paper and illustrated in Figure~\ref{fig:flow_2d_steady}.

\begin{figure}[!ht]
	\centering
	\includegraphics[width=\linewidth]{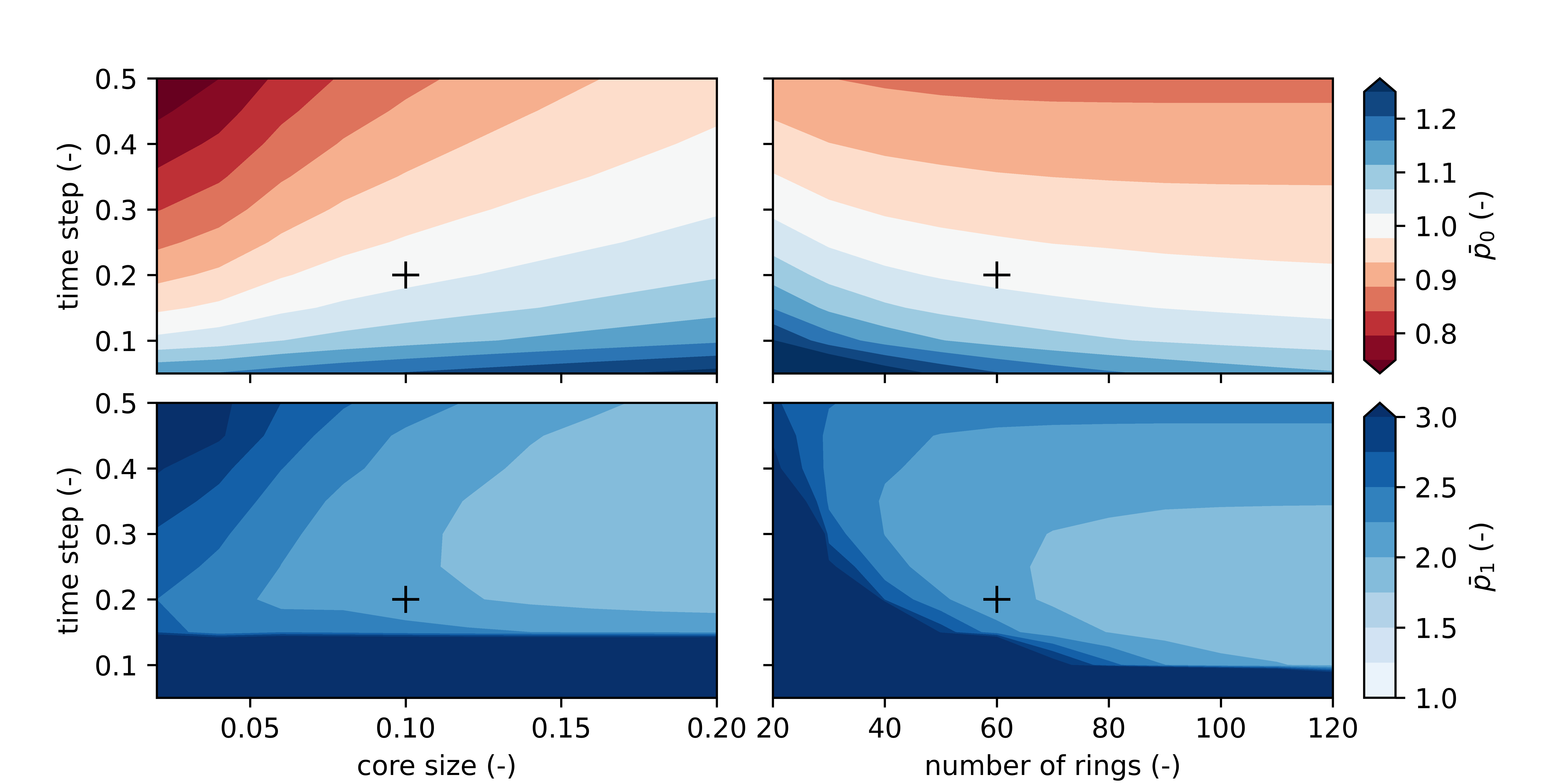}
	\caption{Variation of power for 2D FVW normalised by power estimate from momentum theory, for upstream (top) and downstream (bottom) turbine in a configuration as in Figure~\ref{fig:flow_2d_steady}.
	{The cross indicates the parameter choice used in this paper.}
	The base settings for this parameter sweep are: axial induction $a_0=0.3$ and $a_1=0.33$, number of rings $n_{\mathrm{r}}=60$, time step $h=0.2$, and core size $\sigma=0.1$. }
	\label{fig:parameters_2d}
\end{figure}

The normalised power for a parameter sweep around the operating point from Table~\ref{tab:par_2d} is shown in Figure~\ref{fig:parameters_2d}, evaluating combinations of time step $h$, vortex core size $\sigma$, and number of rings $n_{\mathrm{r}}$.
The chosen parameters appear to be a good match for the normalised power of the upstream turbine, $\bar{p}_0$. 
There is a dependency between time step and core size, so they need to be chosen together. 
The combination of number of rings to model the wake needs to be chosen together with the time-step such that the wake covers the downstream turbine.
As long as the wake adequately covers the downstream turbine position, the power estimate is not very sensitive to variations in the parameters observed here.
The 2D FVW does not exactly represent momentum theory, so it is unsurprising that downstream power is overestimated.

\begin{figure}[!ht]
	\centering
	\includegraphics[width=\linewidth]{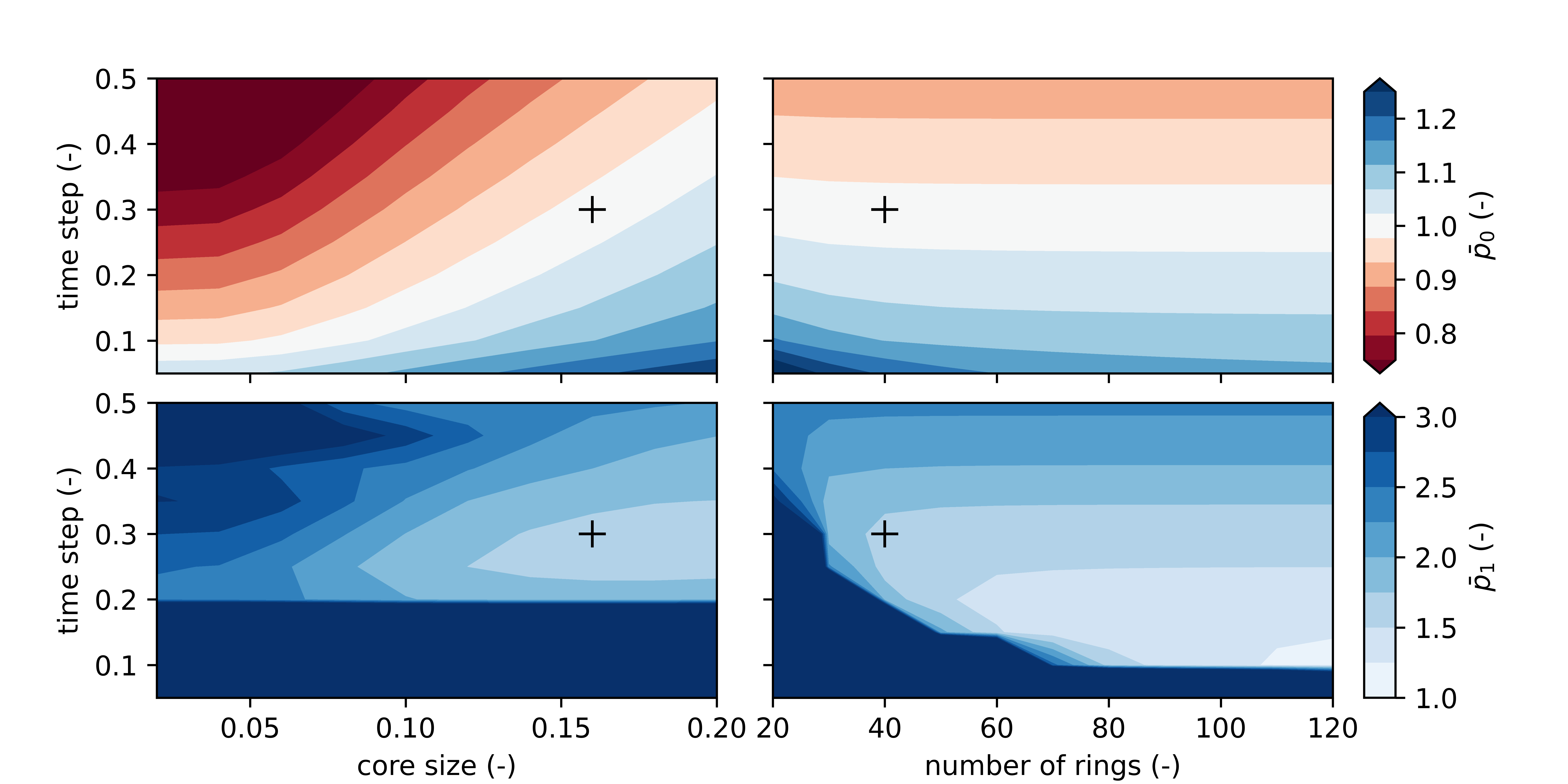}
	\caption{Variation of power for 3D FVW normalised by power estimate from momentum theory, for upstream (top) and downstream (bottom) turbine in a configuration as in Figure~\ref{fig:FVW3D}.
	{The cross indicates  parameter choice used in this paper.}
	{The base settings for this parameter sweep are: axial induction $a_0=0.3$ and $a_1=0.33$, number of rings $n_{\mathrm{r}}=40$, number of elements per ring $n_\mathrm{e}=16$, time step $h=0.3$, and core size $\sigma=0.16$. }
	}
	\label{fig:parameters_3d}
\end{figure}

A similar sweep with the 3D FVW is shown in Figure~\ref{fig:parameters_3d}.
The trends in the results are similar to those shown in the parameter variations for the 2D FVW. 
For this configuration without yaw misalignment, results appeared not very sensitive to the number of elements used in the discretisation of the vortex rings.

\begin{figure}[!t]
	\centering
	\includegraphics[width=0.75\linewidth]{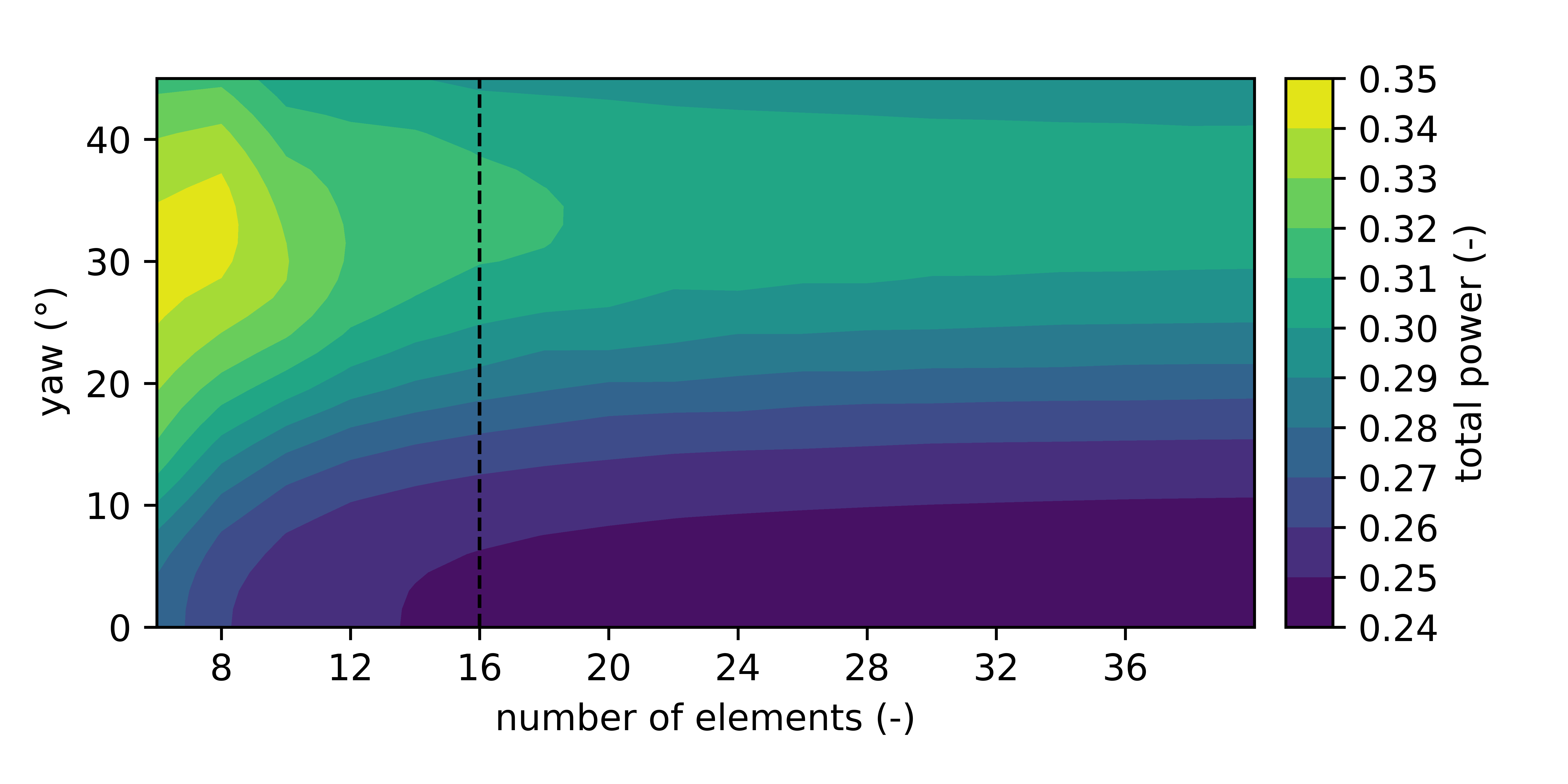}
	\caption{Total power estimate with the 3D FVW under yaw misalignment for varying number of vortex filaments in vortex ring discretisation.
	The dashed line indicates the number of elements used in this paper.
	Configuration as in Figure~\ref{fig:FVW3D}.
	}
	\label{fig:yaw_elements}
\end{figure}

The number of elements in ring discretisation is important for dynamics of wake redirection under yaw misalignment because finer discretisations better represent the curled wake.
Figure~\ref{fig:yaw_elements} shows that power gain from yaw misalignment is observed even for coarse discretisations of the vortex filament rings.
In the trade-off between computational limits and better convergence of the dynamics, a number of elements $n_\mathrm{e}=16$ was chosen for use in this paper.

\section*{Author Contributions}
Maarten J. van den Broek: Conceptualization, Methodology, Software, Validation, Investigation, Writing--Original draft preparation, Visualization.
Delphine De Tavernier: Conceptualization, Methodology, Writing--Review and Editing.
Benjamin Sanderse: Writing--Review and Editing, Supervision.
Jan-Willem van Wingerden: Writing--Review and Editing, Conceptualization, Resources, Project Administration, Funding Acquisition.


\section*{Funding}
This work is part of the research programme ``Robust closed-loop wake steering for large densely spaced wind farms'' with project number 17512, which is (partly) financed by the Dutch Research Council (NWO).

\section*{Data Availability}
The software developed for this study is openly available in figshare with DOI~\href{https:doi.org/10.4121/20278620}{10.4121/20278620} or on GitHub at \url{https://github.com/TUDelft-DataDrivenControl/vortexwake} \citep{vortexcode}.
Additionally, the data presented in this study is openly available in figshare with DOI~\href{https:doi.org/10.4121/20278590}{10.4121/20278590} \citep{vortexdata}.

\bibliographystyle{elsarticle-num-names} 
\bibliography{./wind-farm-control.bib,./references.bib} 





\end{document}